\documentclass{lmcs} 
\usepackage{stmaryrd}


\usepackage{hyperref}

\usepackage{bbm}
\usepackage{listings}

\newtheorem{aside}{Aside}
\newtheorem{example}{Example}
\newtheorem{definition}{Definition}
\newtheorem{theorem}{Theorem}
\newtheorem{proposition}{Proposition}
\newtheorem{lemma}{Lemma}

\makeatletter
\newcommand{\xRightarrow}[2][]{\ext@arrow 0359\Rightarrowfill@{#1}{#2}}
\makeatother

\usepackage{varwidth}
\usepackage{enumitem}

\usepackage{amsmath}
\usepackage{graphicx}
\usepackage{algorithm}
\usepackage{algorithmicx}
\usepackage[noend]{algpseudocode}
\usepackage{color}
\usepackage{xspace}
\usepackage{pgf}

\newcommand{\dom}{\operatorname{\mathbf{dom}}}
\newcommand{\ran}{\operatorname{\mathbf{ran}}}

\usepackage{tikz}
\usetikzlibrary{arrows,snakes,backgrounds,automata,shapes,positioning,chains,calc}

\usepackage[inference]{semantic}

\usepackage{marginfix} 

\newcommand{\OMIT}[1]{}

\usepackage{cutwin}

\theoremstyle{plain} 

\begin{document}

\title[Verifying C11-Style Weak Memory Libraries via Refinement]{Verifying C11-Style Weak Memory Libraries via Refinement}

\author[S.~Dalvandi]{Sadegh Dalvandi}	
\address{University of Surrey, Guildford, Surrey, United Kingdom}	
\email{m.dalvandi@surrey.ac.uk}  
\thanks{Supported by EPSRC Grant EP/R032556/1.}	

\author[B.~Dongol]{Brijesh Dongol}	
\address{University of Surrey, Guildford, Surrey, United Kingdom}	

\email{b.dongol@surrey.ac.uk}  





\newcommand{\linefill}{\cleaders\hbox{$\smash{\mkern-2mu\mathord-\mkern-2mu}$}\hfill\vphantom{\lower1pt\hbox{$\rightarrow$}}}  
\newcommand{\Linefill}{\cleaders\hbox{$\smash{\mkern-2mu\mathord=\mkern-2mu}$}\hfill\vphantom{\hbox{$\Rightarrow$}}}  
\newcommand{\transi}[2]{\mathrel{\lower1pt\hbox{$\mathrel-_{\vphantom{#2}}\mkern-8mu\stackrel{#1}{\linefill_{\vphantom{#2}}}\mkern-11mu\rightarrow_{#2}$}}}
\newcommand{\trans}[1]{\transi{#1}{{}}}
\newcommand{\eseq}[1]{\langle~\rangle}

\newcommand{\transo}[2]{\mathrel{\hbox{$\mathrel=_{\vphantom{#2}}\mkern-8mu\stackrel{#1}{\linefill_{\vphantom{#2}}}\mkern-11mu\Rightarrow_{#2}$}}}

\newcommand{\otrans}[1]{\transo{#1}{{}}}

\newcommand{\traceArrow}[1]{\stackrel{#1}{\longrightarrow}}
\newcommand{\ltsArrow}[1]{\stackrel{#1}{\Longrightarrow}}

\newcounter{sarrow}
\newcommand\strans[1]{%
  \mathrel{\raisebox{0.1em}{
    \stepcounter{sarrow}%
    \!\!\!\!
    \begin{tikzpicture}
      \node[inner sep=.5ex] (\thesarrow) {$\scriptstyle #1$};
      \path[draw,<-,decorate,line width=0.25mm,
      decoration={zigzag,amplitude=0.7pt,segment length=1.2mm,pre=lineto,pre length=4pt}] 
      (\thesarrow.south east) -- (\thesarrow.south west);
    \end{tikzpicture}%
  }}}

\newcommand\Strans[1]{%
\mathrel{\raisebox{0.1em}{
\!\!\begin{tikzpicture}
  \node[inner sep=0.6ex] (a) {$\scriptstyle #1$};
  \path[line width=0.2mm, draw,implies-,double distance between line
  centers=1.5pt,decorate, 
    decoration={zigzag,amplitude=0.7pt,segment length=1.2mm,pre=lineto,
    pre   length=4pt}] 
    (a.south east) -- (a.south west);
\end{tikzpicture}}%
}}

\newcommand{\calE}{{\cal E}}
\newcommand{\nat}{\mathbb{N}}
\newcommand{\noteq}{\neq}

\newcommand{\lt}{{\bf Less than}}

\newcommand{\ev}{\mathit{ev}}
\newcommand{\Events}{\mathit{Evt}}
\newcommand{\Inv}{\mathit{Inv}}
\newcommand{\Resp}{\mathit{Res}}
\newcommand{\his}{\mathit{his}}
\newcommand{\exec}{\mathit{exec}}
\newcommand{\complete}{\mathit{complete}}
\newcommand{\Var}{\mathit{Var}}
\newcommand{\GVar}{{\it GVar}}
\newcommand{\LVar}{\mathit{LVar}}
\newcommand{\Val}{\mathit{Val}}
\newcommand{\Obj}{\mathit{Obj}}
\newcommand{\Meth}{\mathit{Meth}}
\newcommand{\Tid}{\mathit{Tid}}
\newcommand{\CASOp}{\mathit{CAS}} 
\newcommand{\WR}{\mathsf{W_R}}
\newcommand{\RA}{\mathsf{R_A}}
\newcommand{\R}{\mathsf{R}}
\newcommand{\A}{\mathsf{A}}
\newcommand{\RX}{\mathsf{R_X}}
\newcommand{\W}{\mathsf{W}}
\newcommand{\WX}{\mathsf{W_X}}
\newcommand{\CRA}{\mathsf{CRA}}
\newcommand{\URA}{\mathsf{U}}
\newcommand{\URAT}{\mathsf{UT}}
\newcommand{\URAF}{\mathsf{UF}}

\newcommand{\ww}{{\it ww}}
\newcommand{\tts}{{\it tt}}
\newcommand{\mem}{{\it ls}}
\newcommand{\Top}{{\it Top}} 

\newcommand{\HB}{{\sf hb}\xspace} 
\newcommand{\PO}{{\sf po}\xspace}
\newcommand{\MO}{{\sf mo}\xspace}
\newcommand{\SC}{{\sf sc}\xspace}
\newcommand{\RF}{{\sf rf}\xspace}
\newcommand{\SB}{{\sf sb}\xspace}

\newcommand{\refeq}[1]{(\ref{#1})}
\newcommand{\refalg}[1]{Algorithm~\ref{#1}}

\newcommand{\fr}{{\sf fr}}
\newcommand{\ltsb}{{\sf sb}}
\newcommand{\ltrf}{\mathord{\sf rf}}
\newcommand{\ltfr}{{\sf fr}}
\newcommand{\lthb}{{\sf hb}}
\newcommand{\ltsw}{{\sf sw}}
\newcommand{\ltmox}{{\sf mo}^x}
\newcommand{\ltmo}{{\sf mo}}
\newcommand{\lteco}{{\sf eco}}
\newcommand{\PreExec}{{\it PreExec}}
\newcommand{\Approx}{{\it C11}}
\newcommand{\Seq}{{\it Seq}}

\newcommand{\True}{{\it true}}
\newcommand{\False}{{\it false}}

\newcommand{\justified}{justified\xspace}
\newcommand{\notjustified}{unjustified\xspace}
\newcommand{\Justified}{Justified\xspace}
\newcommand{\Notjustified}{Unjustified\xspace}

\newcommand{\rdval}{{\it rdval}}
\newcommand{\wrval}{{\it wrval}}
\newcommand{\loc}{{\it loc}}
\newcommand{\var}{\mathtt{var}}

\newcommand{\imp}{\Rightarrow}
\newcommand{\expr}{\mathit{Exp}}
\newcommand{\flag}{\mathit{flag}}

\newcommand{\hbo}[1]{\stackrel{#1}{\rightarrow}}
\newcommand{\detval}[1]{\stackrel{#1}{=}}
\newcommand{\last}{\mathit{last}}

\newcommand{\dview}{{\it dview}}
\newcommand{\wfs}{{\it wfs}}
\newcommand{\finite}{{\it finite}}
\newcommand{\fval}{\texttt{val}}
\newcommand{\fnxt}{\texttt{nxt}}

\newcommand{\kwnew}{\textsf{\textbf{new}}}

\newcommand{\kwcas}{\textsf{\textbf{CAS}}}
\newcommand{\kwswap}{\textsf{\textbf{swap}}}
\newcommand{\kwskip}{\bot}
\newcommand{\kwdo}{\textsf{\textbf{do}}}
\newcommand{\kwwhile}{\textsf{\textbf{while}}}
\newcommand{\kwend}{\textsf{\textbf{end}}}
\newcommand{\kwif}{\textsf{\textbf{if}}}
\newcommand{\kwthen}{\textsf{\textbf{then}}}
\newcommand{\kwelse}{\textsf{\textbf{else}}}
\newcommand{\kwreturn}{\textsf{\textbf{return}}}
\newcommand{\kwthread}{\textsf{\textbf{thread}}}
\newcommand{\kwuntil}{\textsf{\textbf{until}}}

\newcommand{\whilestep}[1]{\stackrel{#1}{\longrightarrow}}
\newcommand{\fv}{\mathit{fv}}

\algnewcommand\Swap{\kwswap}
\algnewcommand\Skip{\kwskip}
\algnewcommand\Thread{\kwthread}

\algrenewcommand\algorithmicend{\kwend}
\algrenewcommand\algorithmicdo{\kwdo}
\algrenewcommand\algorithmicwhile{\kwwhile}
\algrenewcommand\algorithmicfor{\textsf{\textbf{for}}}
\algrenewcommand\algorithmicforall{\textsf{\textbf{for all}}}
\algrenewcommand\algorithmicloop{\textsf{\textbf{loop}}}
\algrenewcommand\algorithmicrepeat{\textsf{\textbf{repeat}}}
\algrenewcommand\algorithmicuntil{\textsf{\textbf{until}}}
\algrenewcommand\algorithmicprocedure{\textsf{\textbf{procedure}}}
\algrenewcommand\algorithmicfunction{\textsf{\textbf{function}}}
\algrenewcommand\algorithmicif{\kwif}
\algrenewcommand\algorithmicthen{\kwthen}
\algrenewcommand\algorithmicelse{\kwelse}
\algrenewcommand\algorithmicreturn{\kwreturn}

\algblockdefx{Thread}{EndThread}%
[1]{\kwthread \xspace #1}%
{\algorithmicend}

\algblockdefx{MyWhile}{EndMyWhile}%
[1]{\kwwhile \xspace #1}%
{\algorithmicend}

\makeatletter
\ifthenelse{\equal{\ALG@noend}{t}}%
  {\algtext*{EndMyWhile}}
  {}%
\makeatother

\algblockdefx{MyUntil}{EndMyUntil}%
[1]{\kwuntil \xspace #1}%
{\algorithmicend}
\makeatletter
\ifthenelse{\equal{\ALG@noend}{t}}%
  {\algtext*{EndMyUntil}}
  {}%
\makeatother

\makeatletter
\ifthenelse{\equal{\ALG@noend}{t}}%
  {\algtext*{EndThread}}
  {}%
\makeatother

\newcommand{\action}[3]{\ensuremath{
\begin{array}[t]{l@{~}l}
\multicolumn{2}{l}{#1}\\
\textsf{Pre:}&#2\\
\textsf{Eff:}&#3
\end{array}
}}

\newcommand{\lstate}{\mathit{lst}}
\newcommand{\llbr}[1]{\llbracket #1 \rrbracket}
\newcommand{\fresh}{\mathit{fresh}}

\newcommand{\kwtag}{{\it tag}}
\newcommand{\tid}{{\it tid}}
\newcommand{\act}{{\it act}}
\newcommand{\Op}{A}
\newcommand{\Prog}{{\it Prog}}
\newcommand{\Comm}{{\it Com}}
\newcommand{\AComm}{{\it ACom}}
\newcommand{\Exp}{{\it Exp}}
\newcommand{\CExp}{{\it CExp}}
\newcommand{\Init}{\mathbf{Init}}

\newcommand{\initq}{\mathit{qinit}}
\newcommand{\enq}{\mathit{enq}}
\newcommand{\deq}{\mathit{deq}}

\newcommand{\inits}{\mathit{sinit}}
\newcommand{\push}{\mathit{push}}
\newcommand{\pop}{\mathit{pop}}

\newcommand{\Sur}{{\it RC11}}
\newcommand{\asgn}{\ensuremath{:=}}

\newcommand{\bbD}{\mathbb{D}}
\newcommand{\bbE}{\mathbb{E}}
\newcommand{\bbS}{\mathbb{S}}

\newcommand{\rat}{\mathbb{Q}}
\newcommand{\bool}{\mathbb{B}}

\newcommand{\cons}{cons}

\newcommand{\matchedTS}{{\tt matched}}

\newcommand{\maxmo}{\mathbf{max}_{\ltmo}}
\newcommand{\cclose}{\mathbf{cclose}}
\newcommand{\scomp}{\circ}
\newcommand{\view}{\mathit{View}}
\newcommand{\tview}{{\tt tview}}
\newcommand{\ctview}{{\tt ctview}}
\newcommand{\writeson}{{\tt writes\_on}}
\newcommand{\ls}{\mathit{ls}}
\newcommand{\rdview}{\mathit{rdview}}
\newcommand{\isReleasing}{\mathit{isRel}}
\newcommand{\mview}{{\tt mview}}
\newcommand{\mods}{\mathtt{mods}}
\newcommand{\writes}{\mathtt{ops}}
\newcommand{\covered}{\mathtt{cvd}}
\newcommand{\enc}{\mathit{enc}}

\newcommand{\ts}{{\it ts}}

\newcommand{\tst}{{\tt tst}}
\newcommand{\OW}{\mathtt{Obs}}
\newcommand{\visWrites}{\mathit{V\!W}\!}
\newcommand{\encounteredWrites}{\mathit{E\!W}\!}
\newcommand{\Act}{{\sf Act}}

\newcommand{\acquire}{{\it acquire}}
\newcommand{\release}{{\it release}}
\newcommand{\init}{{\it init}}
\newcommand{\maxTS}{{\it maxTS}}

\newcommand{\eqrng}[2]{(\ref{#1}-\ref{#2})}
\newcommand{\refprop}[1]{Proposition~\ref{#1}}
\newcommand{\reffig}[1]{Fig.~\ref{#1}}
\newcommand{\refthm}[1]{Theorem~\ref{#1}}
\newcommand{\reflem}[1]{Lem\-ma~\ref{#1}}
\newcommand{\refcor}[1]{Corollary~\ref{#1}}
\newcommand{\refsec}[1]{Section~\ref{#1}}
\newcommand{\refex}[1]{Example~\ref{#1}}
\newcommand{\refdef}[1]{Definition~\ref{#1}}
\newcommand{\reflst}[1]{Listing~\ref{#1}}
\newcommand{\refchap}[1]{Chapter~\ref{#1}}
\newcommand{\reftab}[1]{Table~\ref{#1}}

\newcommand{\kwfai}{\textsf{\textbf{FAI}}}
\newcommand{\vwrites}{{\tt visible\_writes}}
\newcommand{\wrts}{{\tt writes}}
\newcommand{\isCovered}{\mathit{isCovered}}

\newcommand{\WrX}[2]{#1 := #2}
\newcommand{\WrR}[2]{#1 :=^{\sf R} #2}
\newcommand{\RdX}[2]{#1 \gets #2}
\newcommand{\RdA}[2]{#1 \gets^{\sf A} #2}
\newcommand{\CObs}[5]{\langle #1 = #2\rangle [#4 = {#5}]_{#3}}
\newcommand{\CObss}[3]{\langle #1\rangle [#2]_{#3}}

\newcommand{\COSem}[2]{#1[#2]}
\newcommand{\cvd}[2]{{\bf C}_{#1}^{#2}}
\newcommand{\cvv}[2]{{\bf H}_{#1}^{#2}}

\tikzset{
    mo/.style={dashed,->,>=stealth,thick,black!20!purple},
    hb/.style={solid,->,>=stealth,thick,blue},
    sw/.style={solid,->,>=stealth,thick,black!50!green},
    rf/.style={dashed,->,>=stealth,thick,black!50!green},
    fr/.style={dashed,->,>=stealth,thick,red}
 }

 \lstset{
    mathescape=true,
    breaklines=false,
    tabsize=2,
    morekeywords={if, then, else, let, in },keywordstyle=\color{blue},
    basicstyle=\ttfamily,
    literate={\ \ }{{\ }}1
  }


\begin{abstract}
    Deductive verification of concurrent programs under weak memory
    has thus far been limited to simple programs over a monolithic
    state space.
    For scalability, we also require 
    modular techniques with verifiable library abstractions. This
    paper addresses this challenge in the context of RC11 RAR, a
    subset of the C11 memory model that admits relaxed and
    release-acquire accesses, but disallows, so-called, load-buffering
    cycles. We develop a simple  framework for specifying
    abstract objects that precisely characterises the observability
    guarantees of abstract method calls. We show how this framework
    can be integrated with an operational semantics that enables
    verification of client programs that execute abstract method calls
    from a library they use. Finally, we show how implementations of
    such abstractions in RC11 RAR can be verified by developing a
    (contextual) refinement framework for abstract objects. Our
    framework, including the operational semantics, verification
    technique for client-library programs, and simulation between
    abstract libraries and their implementations, has been mechanised
    in Isabelle/HOL.
\end{abstract}

\maketitle


\section{Introduction}

The formal verification of concurrent programs in a weak memory setting
is challenging since various writes to the memory locations
may be reordered arbitrarily and therefore seen by other threads
in an order different from the program order. In order to reason 
about such programs, the
\emph{observations} that a thread can make of the writes within
the system should be taken into consideration.
In recent work, reasoning about per-thread observations has led to operational
characterisations of the memory model, where high-level predicates for
reasoning about per-thread observations have been developed, and deductive verification
techniques applied to standard litmus tests and non-trivial synchronisation
algorithms~\cite{ECOOP20}.





Current operational verification techniques for weak memory are however, focussed on (closed)
programs, and hence 
do not provide straightforward mechanisms for (de)composing clients
and libraries. This problem requires special consideration under weak
memory since the execution of a library
method 
may induce synchronisation. That is, a thread's observations of a system
(including of client variables) can change 
when executing 
library methods.

This paper addresses several questions surrounding client-library
composition in a weak memory context. Like prior works
\cite{ECOOP20,DBLP:conf/ppopp/DohertyDWD19,DBLP:conf/ecoop/KaiserDDLV17},
we focus on RC11-RAR, a fragment of C11 memory model, which is one of
the most interesting and challenging fragments for
verification.\footnote{The memory model can be extended, but since
  this paper focusses on verification and refinement we keep the
  memory fragment small to simplify the presentation. }

{\bf (1)} {\em How can a client \emph{use} a weak memory library,
i.e., what abstract guarantees can a library provide a client program?}
Prior works~\cite{DongolJRA18,DBLP:conf/popl/BattyDG13} describe
techniques for \emph{specifying} the behaviour of abstract objects,
which are in turn related to their implementations using causal
relaxations of linearizability. 
However, these works do not provide a mechanism for reasoning about
the behaviour of client programs that {\em use} abstract libraries. In
this paper, we address this gap by presenting a modular operational
semantics 
that combines weak
memory states of clients and libraries.
This is our first
contribution.

{\bf (2)} {\em What does it mean to \emph{implement} an abstract
  library?}  To ensure that behaviours of client programs using an
abstract library are preserved, we require \emph{contextual
  refinement} between a library implementation and its abstract
specification. This guarantees that no new client behaviours are
introduced when a client uses a (concrete) library implementation in
place of its (abstract) library specification. Under sequential
consistency (SC), it is well known that linearizable libraries
guarantee (contextual)
refinement~\cite{DBLP:conf/icfem/DongolG16,GotsmanY11,DBLP:journals/tcs/FilipovicORY10}. However,
under weak memory, a generic notion of linearizability is difficult to
pin down~\cite{DongolJRA18,ifm18}. We therefore present a direct
technique for establishing contextual refinement under weak memory. A
key innovation is the development of context-sensitive simulation
rules that ensure that each client thread that uses the
implementation observes a subset of the values seen by the
abstraction. This is our second contribution.


{\bf (3)} {\em Can the same abstract library specify \emph{multiple}
  implementations?}  A key benefit of refinement is the ability to use
the same abstract specification for multiple implementations, e.g., to
fine-tune clients for different concurrent workload scenarios. To
demonstrate the applicability of our framework, we provide a
proof-of-concept example for an abstract lock and show that the same
lock specification can be implemented by a sequence lock and ticket
lock. The theory itself is generic and can be applied to concurrent
objects in general; we prove correctness of the Treiber
Stack for RC11-RAR. The original Treiber stack assumes SC memory. We prove correctness under relaxed and release-acquire accesses, thereby showing where additional synchronisation is required under weak memory. This is our third contribution.

{\bf (4)} {\em How can we support operational verification? Can the
  verification techniques be mechanised?}
Assuming the existence of an operational semantics for the underlying
memory model, we aim for \emph{deductive} verification of both
client-library composition and contextual refinement. We show that
this can be supported 
by prototyping the full verification stack in the Isabelle/HOL theorem
prover.\footnote{Our Isabelle theories may be accessed via~\cite{Isabelle}. Note that the theories require Isabelle 2020.} All of the programs, theorems and proof rules in this paper are fully verified. Our Isabelle development is around 25 KLOC (thousand lines of code), where around 2.5 KLOC is taken from~\cite{DBLP:journals/darts/DalvandiDDW20}.
This is our fourth contribution.

Closely related to our work is that of Raad et al. \cite{DBLP:journals/pacmpl/RaadDRLV19}, 
who have tackled the
problem of client-library programs in a declarative
framework. Their framework provides support for axiomatic 
specifications of weak memory concurrent libraries and verification of
library implementations against these specifications. Although this work addresses a similar set of questions to ours, there are several key differences in our approach.
We move from a declarative to an operational semantics, which provides
a significant step-change as it allows integration with
well-understood reasoning techniques of Owicki-Gries
(cf. \refsec{sec:hoare-logic-c11}) and refinement
(cf. \refsec{sec:cont-refin}), as well as the possibility of
rely-guarantee in future work. Raad et al
\cite{DBLP:journals/pacmpl/RaadDRLV19} also describe a notion of
refinement, but their definition is not based on trace
refinement. Moreover, our operational semantics enables a new approach
to library specifications than Raad et
al. \cite{DBLP:journals/pacmpl/RaadDRLV19}. In particular, we follow
an approach in which library operations interact with weak memory in a
manner that is similar to the interaction between standard reads and
writes within C11 (see \refsec{sec:specifying-c11-style}). This paves
the way for implementations that take further advantage of weak
memory, since our specifications allow new operations to be introduced
in the ``middle'' of an existing abstract history. This is akin to
checking multiple complete executions in a declarative approach for each legal position of the operation.  

The strength of Raad et al.'s declarative approach
\cite{DBLP:journals/pacmpl/RaadDRLV19} is that it considers more than
one memory model and handles so-called ``towers of abstraction'',
where a library implementation can in turn use other library
specifications. The former is more complex in an operational setting
since the operational rules are fine-tuned to a particular memory
model, but the latter can be handled via straightforward extensions to
the semantics (see Aside~\ref{aside:1}).\footnote{We have not done
  this here to simplify the presentation.} On the other hand, our approach
  is in a stronger position w.r.t mechanised verification. While  our 
  development (including all examples and case studeis) is fully mechanised,
  \cite{DBLP:journals/pacmpl/RaadDRLV19} has only achieved a partial
  mechanisation over a small parts of their contribution.

    \smallskip
\noindent{\bf Contributions.}
Overally, the contributions of our work  can be seen as follows.
 First, it presents an
approach for specifying abstract libraries and the associated view-based
assertions that allows verification of client programs that use the 
abstract library in an Owicki-Gries setting. Second, it introduces 
a generalised  operational semantics for C11 weak memory that allows
implementation of C11-style libraries and also provides an assertion 
language that allows for verification of libraries at the implementation
level. Third, it provides a refinement
framework that allows one to prove that an implementation correctly 
refines the abstract specification. The first two, can be used independently
and offers enough support (in varying levels of abstraction) to verifying
client-library programs. The third contribution, allows for establishing
a formal link between the high level abstract specification of a library
and its implemenation.

 The general workflow of our approach is as follows:
\begin{itemize}
  \item Defining the abstract operational specification of a library.
  \item Verifying client programs that uses the library operations using Hoare logic
  and Owicki-Gries.
  \item Implementing the library using the generalised operational 
  semantics of C11.
  \item Proving that the implementation is a contextual refinement of the
  abstract specification and therefore it can safely replace the abstract
  specification in the client programs.
\end{itemize}

  \smallskip
\noindent{\bf Overview.}
In \refsec{sec:message-passing-via} we motivate the main ideas and
illustrate the basic principle of message-passing synchronisation, but
for programs comprising a program executing with an abstract
library. \refsec{sec:gener-weak-memory} presents a modular programming language
and operational semantics for reads/write accesses within this
language. A method for encoding the operational semantics of abstract
objects is given in \refsec{sec:abstr-object-semant}. In
\refsec{sec:example-verification}, we describe verification of client
programs that use abstract libraries, and in \refsec{sec:cont-refin}
we present our refinement theory. We use an abstract lock and two
different implementations of this lock as running examples. 
Finally, in \refsec{sec:view-based-relax}, we discuss the
specification of generic concurrent objects that exploit view-based
relaxations, making use of the framework developed in
\refsec{sec:abstr-object-semant}. Here, we also describe the
verification of the Treiber Stack adapted for C11.


\section{Message passing via library objects}
\label{sec:message-passing-via}



We first illustrate the basic principles of client-object synchronisation in weak memory. 

\begin{figure}[t]
  \begin{minipage}[b]{0.46\columnwidth}
      \begin{center} 
  {\bf Init: } $d:=0;$ $s.init();$  \\
  $\begin{array}{@{}l@{\ }||@{\ }l}
     \text{\bf Thread } 1
     & \text{\bf Thread } 2\\
     d := 5; \qquad & 
                       \text{\bf do } r_1 := s.pop() \\
          
     s.push(1); & \text{\bf until}\ r_1 = 1;  \\ 
     & r_2 \gets d; \\
     \end{array}$

   {\color{red!70!black} $\{r_2 = 0 \lor r_2=5\}$}  \qquad \quad  \quad     
 \end{center}
 \caption{Unsynchronised message passing}
 \label{fig:po-message-bad}
\end{minipage}
\hfill
  \begin{minipage}[b]{0.51\columnwidth}
  \begin{center}  
  {\bf Init:} $d:=0;$ $s.init();$ \\
  $\begin{array}{@{}l@{\ }||@{\ }l}
     \text{\bf Thread } 1
     & \text{\bf Thread } 2\\
     d := 5; \qquad & 
                      \text{\bf do } r_1 := s.pop^{\sf A}() \\
     s.push^{\sf R}(1);               & \text{\bf until}\ r_1 = 1;  \\ 
     
     & r_2 \gets d; \\
 
     \end{array}$

   {\color{green!40!black} $\{ r_2=5\}$}  \qquad   \  \ \ 
 \end{center}
 \caption{Publication via a synchronising
   stack 
 }
 \label{fig:publication}
\end{minipage}
 \end{figure}

\smallskip \noindent{\bf Client-object message passing.}
Under SC all threads have a single common \emph{view} of the shared
state. When a new write is executed, the ``views'' of all threads are
updated so that they are guaranteed to only see this new latest write. In
contrast, each thread in a C11 program has its own view of each
variable. Views may not be updated when a write occurs, allowing
threads to read stale writes. To enforce view updates, additional
synchronisation (e.g., release-acquire) must be introduced~\cite{DBLP:conf/popl/BattyOSSW11,DBLP:conf/ecoop/KaiserDDLV17,DBLP:conf/pldi/LahavVKHD17}.



Now consider a generalisation of this idea to (client) programs that use
library
objects. 
The essence of the problem is illustrated by the message-passing programs in Figures~\ref{fig:po-message-bad}~and~\ref{fig:publication}. 
Under SC, when the program in
\reffig{fig:po-message-bad} terminates, the value of $r_2$ is
guaranteed to be $5$. 
However, this is not necessarily true in a weak memory setting. Even
if $pop$ operation in thread~2 returns 1, it may be possible for
thread 2 to observe a stale value 0 for $d$. Therefore the program only
guarantees the weaker postcondition $r_2 = 0 \lor r_2 = 5$.

To address this problem, the library operations in
\reffig{fig:publication} are annotated with release-acquire
annotations. In particular, the client assumes the availability of a
``releasing push'' ($push^{\sf R}(1)$), which is to be used for
message passing. Thread~2 pops from $s$ using an ``acquiring pop''
($pop^{\sf A}()$). If this pop returns 1, the stack operations induce
a happens-before synchronisation in the client, which in turn means
that it is now impossible for thread 2 to read the stale initial
write for $d$.




\smallskip\noindent{\bf Verification strategy.}
Our aim is to enable {\em deductive verification} of such programs by
leveraging recently developed operational semantics, assertion
language and Owicki-Gries style proof strategy for RC11
RAR~\cite{ECOOP20}. We show that these existing concepts generalise
naturally to client-object, and in a manner that enables modular
proofs.

\begin{figure}[t]
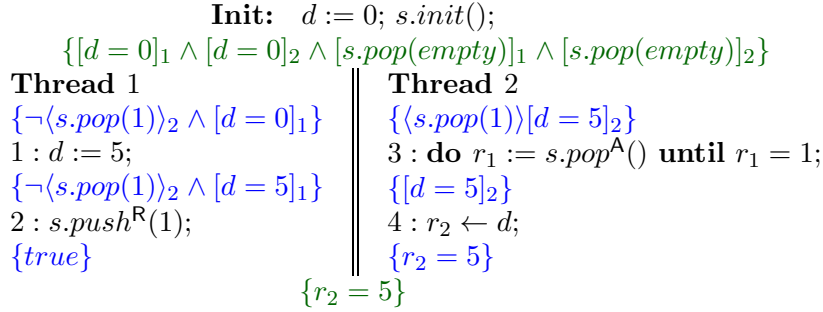

  \centering
 

\begin{minipage}[t]{0.9\columnwidth}
    \begin{center} 
  {\bf Init: } $d:=0;$ $s.init();$  \qquad \qquad \qquad \qquad\\ 
   {\color{green!40!black} $\{[d = 0]_1 \wedge [d = 0]_2 \wedge [s.pop({empty})]_1 \wedge [s.pop({empty})]_2\}$} \\ 
  $\begin{array}{l@{\quad}||@{\quad }l}
     \text{\bf Thread } 1
     & \text{\bf Thread } 2\\
     \begin{array}[t]{@{}l@{}}
       {\color{blue} \{ \neg \langle s.pop(1) \rangle_2 \wedge [d = 0]_1\}} \\
       1: d := 5; \\
       {\color{blue} \{ \neg \langle s.pop(1) \rangle_2 \wedge [d = 5]_1\}} \\
       2: s.push^{\sf R}(1); \\
       {\color{blue} \{ true\}}
     \end{array}
     & 
       \begin{array}[t]{@{}l@{}}
         {\color{blue} \{ \langle s.pop(1) \rangle [d= 5]_2 \}} \\ 
         3: \text{\bf do } r_1 := s.pop^{\sf A}()\ 
         \text{\bf   until}\ r_1 = 1; \\
         {\color{blue} \{ [d = 5]_2\}} \\
         4: r_2 \gets d; \\
         {\color{blue} \{r_2 = 5\} }
     \end{array}
   \end{array}
   $

   {\color{green!40!black} $\{ r_2=5\}$} \qquad {}\qquad {}\qquad {}
 \end{center}
  \caption{A proof outline for message passing}
  \label{fig:mp-proof}

\end{minipage}

\end{figure}

The assertion language of~\cite{ECOOP20} enables reasoning about a
thread's views, e.g., in \reffig{fig:mp-proof}, after initialisation,
thread $t \in \{1,2\}$ has \textit{definite value} $0$ for $d$
(denoted $[d = 0]_t$).
In this paper, we extend such assertions to capture thread views over
library objects. E.g., after initialisation, the only value a pop by
thread $t$ can return is $empty$, and this is captured by the
assertion $[s.pop({empty})]_t$. 
The precondition of $d := 5$ states that thread 2 cannot pop value
$1$ from $s$ (as captured by the assertion $\neg \langle
s.pop(1)\rangle_2$).
The precondition of the $\kwuntil$ loop in thread 2 contains a {\em
  conditional observation} assertion (i.e.,
$\langle s.pop(1) \rangle [d = 5]_2$), which states that if thread~2
pops value 1 from $s$ then it will subsequently be in a state where it
will definitely read $5$ for $d$.

A key benefit of the logic in~\cite{ECOOP20} is that it enables the use of
{\em standard} Owicki-Gries reasoning and straightforward
mechanisation~\cite{DBLP:journals/corr/abs-2004-02983}. As we shall see (\refsec{sec:example-client-lbjec}), we maintain these benefits in
the context of client-object programs.



\smallskip\noindent{\bf Contextual refinement.}
\label{sec:cont-refin-pre}
Contextual refinement relates a client using an abstract object with a
client that uses a concurrent implementation of the object. More precisely, we say
that a concrete object $CO$ is a \emph{contextual refinement} of an
abstract object $AO$ iff for any client $C$, every behaviour of $C$
when it uses $CO$ is a possible behaviour of $C$ when it uses
$AO$. Thus, there is no observable difference to any client when it
uses $CO$ in place of $AO$.

In a weak memory setting, to enable a client to \emph{use} an object,
one must \emph{specify} how synchronisation between object method
calls affects the client state. To {\em implement} such a
specification, we must describe how the abstract synchronisation
guarantees 
are represented in the implementation. Prior works have appealed to
extensions of notions such as linearizability to ensure contextual
guarantees
\cite{DongolJRA18,DBLP:journals/pacmpl/EmmiE19,DBLP:journals/pacmpl/RaadDRLV19}. In
this paper, we aim for a more direct approach and consider contextual
refinement 
 directly. 



\section{Generalised 
  operational semantics}
\label{sec:gener-weak-memory}
We now present a simple program syntax 
that allows one to write open programs 
that can be filled by an abstract method or concrete implementation of
a method.


\subsection{Program Syntax}
\label{sec:program-syntax}

We start by defining a syntax of concurrent programs, starting with
the structure of sequential programs (single threads).  A thread may
use {\em global} shared variables (from $\GVar$) and local registers
(from $\LVar$).  We let $\Var = \GVar \cup \LVar$ and assume
$\GVar \cap \LVar = \emptyset$. For client-library programs, we 
partition $\GVar$ into $\GVar_C$ (the global client variables) and
$\GVar_L$ (the global library variables) and similarly $\LVar$ into
$\LVar_C$ and $\LVar_L$. In an implementation, global variables can be
accessed in three different {\em synchronisation modes}: acquire ({\sf
  A}, for reads), release ({\sf R}, for writes) and relaxed (no
annotation).  The annotation {\sf RA} is employed for {\em update}
operations, which read and write to a shared variable in a single
atomic step.  
We let $\Obj$ and $\Meth$ be the set of all objects and method calls,
respectively.

We assume that
$\ominus$ is a unary operator (e.g., $\neg$), $\oplus$ is a binary
operator (e.g., $\land$, $+$, $=$) and $n$ is a value (of type
$\Val$). Expressions must only involve local
variables. 
The syntax of sequential programs, $\Comm$, is given by the following
grammar with $r \in \LVar, x \in \GVar, o \in \Obj, m \in \Meth, u, v \in \Val$:

\bigskip
\noindent
\begin{tabular}[t]{r@{~}ll}
  $\Exp_L$   ::= & $\Val \mid \LVar  \mid \ominus \Exp_L \mid \Exp_L \oplus Exp_L$ 
  \\
  $\CExp_L$   ::= & $\bullet \mid \Exp_L$ \\[1mm]
  $\bullet$ ::= & $\Val \mid o.m([u]) \mid \Comm $, where $\Comm$ contains no holes \\[1mm]
  
  $\AComm$ ::= & $ \bullet \mid \kwskip \mid 
                 r \gets \kwcas^{\sf [R] [A]}(x, u, v) 
\mid$ \\
              &$r \gets \kwfai^{\sf [R] [A]}(x)
                   \mid r := \CExp_L  
                 \mid {} $
   $x :=^{\sf [R]} \Exp_L \mid  r \gets^{\sf [A]} x$ \\[1mm] 
  $\Comm$ ::= & $\AComm \mid \Comm ; \Comm \mid \kwif~B\ \kwthen\ \Comm\ \kwelse\ \Comm \mid  
                    \kwwhile\ B\ \kwdo\ \Comm$
\end{tabular} \bigskip

\noindent where we assume $B$ to be an expression of type $\CExp_L$
that evaluates to a boolean. 
We allow programs with holes, denoted
$\bullet$, which may be filled by an abstract or concrete method
call. During a program's execution, the hole may also be filled by
the null value $\bot \notin
\Val$, or the return value of the method call.
The notation ${\sf [X]}$ denotes that the annotation ${\sf X}$ is
optional, where ${\sf X} \in \{{\sf A}, {\sf R}\}$, enabling one to
distinguish relaxed, acquiring and releasing accesses. Within a method
call, the argument $u$ is optional. Later, we will also use
$\kwdo$-$\kwuntil$ loops, which is straightforward to define in terms
of the  syntax above.

\subsection{Program Semantics}

\label{sec:program-semantics}

For simplicity, we assume concurrency at the top level only. We let
$\Tid$ be the set of all thread identifiers and use a function
$\Prog: \Tid \to \Comm$ to model a program comprising multiple
threads. In examples, we typically write concurrent programs as
$C_1 || \ldots || C_n$, where $C_i \in \Comm$.  We further assume some
initialisation of variables.  The structure of our programs
thus is $\Init; \big( C_1 || \ldots || C_n \big) $.

The operational semantics for this language is defined in three parts.
The \emph{program semantics} fixes the steps that the concurrent
program can take.  This gives rise to transitions
$(P,\lstate) \trans {a}_t (P',\lstate')$ of a thread $t$ where $P$ and
$P'$ are programs, $\lstate$ and $\lstate'$ is the state of local
variables and $a$ is an action (possibly the silent action $\epsilon$,
see below).  The program semantics is combined with a {\em memory
  semantics} reflecting the C11 state, and in particular the writes from which a read action can
read. 
Finally, there is the \emph{object semantics}, which defines the
abstract semantics of the object at hand.


We assume that the set of actions is given by $\Act$. We let
$\epsilon \notin \Act$ be a silent action and let
$\Act_\epsilon = \Act \cup \{\epsilon\}$. 
For an
expression $E$ over local variables, we write $\llbr{E}_{\ls}$ for the
value of $E$ in local state $\ls$; we write $\ls[r := v]$ to state
that $\ls$ remains unchanged except for the value of local variable
$r$ which becomes $v$. We use $C[D]$ to denote the program $C$ with the 
hole filled by $D$. If $D = \bot$, we proceed with the execution of
$C$, otherwise we execute $D$. Note that if $D$ terminates with a
value (due to a method call that returns a value), then the hole
contains a value and execution may proceed by either using the rule
for $r \asgn v$ or the rule for $v ; C_2$, both of which are present
in \reffig{fig:comm-sem}. The last two rules, {\sc Cli} and {\sc Lib},
lift the transitions of threads to a transition of a client and
library program, respectively. These are distinguished by the
subscript $L$, which only appears in transitions corresponding to the
library.

\begin{figure*}[t]\small
  \centering %
  $\inference{r \in \LVar \quad v = \llbr{E}_{{\it ls}}}{(r := E,{\it ls}) \trans{\epsilon} (\kwskip,\ls[r := v]) }\hfill
    \inference{x \in \GVar \quad a = wr^{\sf [R]}(x, \llbr{E}_{{\it ls}}) }{(x :=^{\sf [R]} E,ls) \trans{a} (\kwskip,\ls) }\hfill \inference{a = rd^{\sf [A]}(x,v) \quad v \in \Val}{ (r \gets^{\sf [A]} x, \ls) \trans a (\kwskip,\ls[r := v]) }
  $
   \bigskip
  
$
   \inference{(C_1,\ls) \trans{a} (C_1',ls')}{(C_1 ; C_2,\ls) \trans{a}
    (C_1' ; C_2,\ls')}\hfill
     \inference{v \in \Val \cup \{\bot\}}{(v ; C_2,\ls) \trans{\epsilon}
    (C_2,\ls)}\hfill
    \inference
  {\llbr{B}_{\ls}}
  {({\it IF}, \ls) \trans{\epsilon} (C_1,\ls)} 
    \hfill\inference
  {\neg \llbr{B}_{\ls}}
  {({\it IF}, \ls) \trans{\epsilon} (C_2,\ls) }
  $
 \bigskip

 $
    \inference{\llbr{B}_{\ls}}{
    \begin{array}[t]{@{}l@{}}
      ({\it WHILE}, \ls) 
      \trans{\epsilon}  (C; {\it WHILE}, \ls) 
    \end{array}
  }     \qquad
  \inference{\neg \llbr{B}_{\ls}}
  {
    \begin{array}[t]{@{}l@{}}
      ({\it WHILE}, \ls)  \trans{\epsilon}  (\kwskip, \ls) 
    \end{array}
  }    \qquad
\inference{}{ (\COSem{C}{\kwskip}, \ls) \trans \epsilon (C,\ls) }$
  
\bigskip
      $  \inference{(D, ls) \trans a (D', ls')}{ (\COSem{C}{D}, \ls) \trans{a}_L (\COSem{C}{D'},\ls') }
       \qquad
   \inference{a = rd(x,v') \quad v'\neq u \quad u,v,v' \in \Val}{ (r \gets \kwcas^{\sf [R] [A]}(x, u, v), \ls) \trans a (\kwskip,\ls[r := \False]) }
$
\bigskip

  $
    \inference{a = upd^{\sf X}(x,u,v) \quad u,v \in \Val}{ (r \gets \kwcas^{\sf X}(x, u, v), \ls) \trans a (\kwskip,\ls[r := \True]) }
    \qquad
    \inference{a = upd^{\sf X}(x,u,u+1) \quad u \in \Val}{ (r \gets \kwfai^{\sf X}(x), \ls) \trans a (\kwskip,\ls[r := u]) }
  $ \bigskip

$
  \inference[\sc
  Cli]{(P(t), \ls) \trans{a} (C,\ls') \quad a \in \Act_\epsilon}
  {(P,\lstate) \trans{a}_t (P[t := C], \ls')}
  \qquad
  \inference[\sc
  Lib]{(P(t),\lstate) \trans{a}_L (C,\ls') \quad a \in \Act_\epsilon}
  {(P,\lstate) \trans{a}_{L, t} (P[t := C], \ls')}
  $
  
\caption{Program semantics, where ${\it IF} = \kwif \ B\ \kwthen\
    C_1\ \kwelse\ C_2$ and ${\it WHILE} = \kwwhile\ B\ \kwdo\ C$}
  \label{fig:comm-sem}

\end{figure*}

The rules in \reffig{fig:comm-sem} allow for {\em all} possible values
for any read. We constrain these values 
with respect to a {\em memory semantics} (formalised by
$\strans{a}_t$), which is described for reads, writes and updates in
\refsec{sec:memory-semantics} and for abstract objects in
\refsec{sec:abstr-object-semant}.
The combined semantics brings together a client state $\gamma$ and library
state $\beta$ as follows.
\begin{gather*} \small
  \inference{(P,\lstate) \trans {\epsilon }_t (P',\lstate')}
  {(P,\lstate,\gamma, \beta) \ltsArrow{
    } (P', \lstate',\gamma, \beta)}  \small
  \qquad  
  \inference{(P,\lstate) \trans {\epsilon }_{L,t} (P',\lstate')}
  {(P,\lstate,\gamma, \beta) \ltsArrow{
    }
    (P', \lstate',\gamma, \beta)}
\\[1mm] \small
  \inference{(P,\lstate) \trans {a}_t (P',\lstate') \qquad  \gamma,
    \beta \strans{a}_{t} \gamma', \beta'} {(P,\lstate,\gamma, \beta)
    \ltsArrow{
    } (P', \lstate',\gamma', \beta')} 
  \qquad 
  \inference{(P,\lstate) \trans {a}_{L,t} (P',\lstate') \qquad 
    \beta, \gamma \strans{a}_{t} \beta', \gamma'}
  {(P,\lstate,\gamma, \beta)
    \ltsArrow{
    } (P', \lstate',\gamma', \beta')}
\end{gather*}
These rules ensure, for example, that a read only returns a value
allowed by the underlying memory model. In
\refsec{sec:abstr-object-semant}, we introduce additional rules so
that the memory model also contains actions corresponding to method
calls on an abstract object.

Note that the memory semantics (see \refsec{sec:memory-semantics} and \refsec{sec:abstr-object-semant}) defined
by $\gamma, \beta \strans{a}_t \gamma', \beta'$ assumes that $\gamma$ is
the state of the component being executed and $\beta$ is the state of
the context. For a client step, we have that $\gamma$ is the
executing component state and $\beta$ is the context state, whereas
for a library step, these parameters are swapped.

\subsection{Memory Semantics}
\label{sec:memory-semantics}

Next, we detail the modularised memory semantics, which builds on an
earlier monolithic semantics~\cite{ECOOP20}, which is a
timestamp-based revision of an earlier operational
semantics~\cite{DBLP:conf/ppopp/DohertyDWD19}. Our present extension
is a semantics that copes with client-library interactions in weak
memory. Namely, it describes how synchronisation (in our example
release-acquire synchronisation) in one component affects thread views
in another component. The semantics 
accommodates both client synchronisation affecting a library, and vice
versa.

\smallskip\noindent {\bf Component State.} 
We assume $\Act$ denotes the set of actions. Following~\cite{ECOOP20},
each global write is represented by a pair
$(a, q) \in \Act \times \rat$, where $a$ is a write action, and $q$ is
a rational number that we use as a {\em timestamp} corresponding to
modification order (cf.
\cite{DBLP:conf/ecoop/KaiserDDLV17,Dolan:2018:LDRF,DBLP:journals/corr/PodkopaevSN16}).
The set of modifying operations within a component that have occurred
so far is recorded in $\writes \subseteq \Act \times \rat$. Unlike
prior works, to accommodate (abstract) method
calls of a data structure, we record abstract operations in general,
as opposed to writes only.

Each state must record the operations that are observable to each
thread. To achieve this, we use two families of functions from global
variables to writes~(cf.
\cite{DBLP:journals/corr/PodkopaevSN16,DBLP:conf/popl/KangHLVD17}). 
\begin{itemize}[leftmargin=*]
\item A \emph{thread view} function $\tview_t \in \GVar \rightarrow \writes$ that returns the
  \emph{viewfront} of thread $t$. The thread $t$ can read from any
  write to variable $x$ whose timestamp is not earlier than
  $\tview_t(x)$. Accordingly, we define, for each state $\gamma$,
  thread $t$ and global variable $x$, the set of {\em observable
    writes}, where $\tst(w) = q$ denotes $w$'s timestamp:

  \vspace{2pt}\hfill$
    \gamma.\OW(t, x) =  \{(a, q) \in \gamma.\writes  \mid 
                                     \begin{array}[t]{@{}l@{}}
\mathit{var}(a) = x    {}\wedge \tst(\gamma.\tview_t(x)) \leq q\}
                                     \end{array}
  $ \hfill{} \vspace{2pt}

\item A \emph{modification view} function $\mview_w \in  \GVar \rightarrow \Act \times \rat$ that records the
  \emph{viewfront} of write $w$, i.e., the viewfront of the thread
  that executed $w$ immediately after $w$'s execution. We use
  $\mview_w$ to compute a new value for $\tview_t$ if a thread $t$
  \emph{synchronizes} with $w$, i.e., if $w \in \WR$ and another
  thread executes an $e \in \RA$ that reads from $w$.
\end{itemize}
The client cannot directly access writes in the library, therefore the thread
view function must map to writes within the same component. On the other
hand, synchronisation in a component can affect thread views in
another (as discussed in \refsec{sec:message-passing-via}), thus the
modification view function may map to operations across the system.


Finally, our semantics maintains a set $\covered \subseteq
\writes$. In C11 RAR, each update action occurs in modification order
immediately after the write that it reads from
\cite{DBLP:conf/ppopp/DohertyDWD19}. This property ensures the
atomicity of updates. 
We disallow 
any newer modifying operation (write or update) from intervening
between any update and the write or update that it reads from.  As we
explain below, covered writes are those that are immediately prior to
an update in modification order, and new write actions never interact 
with a covered write (i.e. do not write immediately after them in the modification order).

\smallskip
\noindent{\bf Initialisation.} 
Suppose
$\GVar_C = \{x_1, ... , x_n\}$, $\GVar_L = \{y_1, ..., y_{n'}\}$, 
$\LVar = \{r_1, ... , r_m\}$, 
$k_i, k_i', l_i \in \Val$, and
$\Init = x_1:=k_1; \ldots, x_n:=k_n ; [r_1 := l_1 ;] \dots [r_m :=
l_m;]$, where we use the notation $[r_i := l_i;]$ to mean that the
assignment $r_i := l_i$ may optionally appear in $\Init$. Thus each
shared variable is initialised exactly once and each local variable is
initialised at most once. The initial values of the state components
are then as follows, where we assume $0$ is the initial
timestamp, $t$ is a thread, $x_i \in \GVar_C$ and $y_i \in \GVar_L$
\begin{align*}
  & \gamma_\Init.\writes  = \bigcup_{1 \leq i \leq n }\{(wr(x_i,k_i),0)\} 
   & 
  & \beta_\Init.\writes  = \bigcup_{1 \leq i \leq n' }\{(wr(y_i,k_i'),0)\} \\
  & \gamma_\Init.\tview_t(x_i) =  (wr(x_i,k_i),0)
   & &
    \beta_\Init.\tview_t(y_i) =  (wr(y_i,k_i'),0) \\ 
  & \gamma_\Init.\covered  = \beta_\Init.\covered= \emptyset & & 
  \gamma_\Init.\mview_{x_i} =
   \begin{array}[t]{@{}l@{}}
     \beta_\Init.\mview_{y_i}= \\\gamma_\Init.\tview_t \!\cup\! \beta_\Init.\tview_t
                                                                 \end{array}
\end{align*}

The local state component of each thread must also be compatible with
$\Init$, i.e., for each $t$ if $r_i \in \dom(lst(t))$ we have that
$({\it lst}(t))(r_i) = l_i$ provided $r_i := l_i$ appears in $\Init$.
We let ${\it lst}_\Init$ be the local state compatible with $\Init$
and let $\Gamma_\Init = ({\it lst}_\Init,\gamma_\Init, \beta_\Init)$.


\begin{figure*}[t]
  \centering \small
  $\inference[{\sc Read}] {a \in \{rd(x, n), rd^\mathsf{A}(x, n) \}
    \qquad (w, q) \in \gamma.\OW(t, x) \qquad
    \wrval(w) = n \\
    \tview' = \ensuremath{
      \begin{cases}
        \gamma.\tview_t\otimes\gamma.\mview_{(w,q)}&\mbox{if
          $(w, a) \in \WR \times \RA$ 
         }\\
         \gamma.\tview_t[x := (w, q)]&\mbox{otherwise}
       \end{cases}} 
     \\    \ctview' = \ensuremath{
      \begin{cases}
        \beta.\tview_t\otimes\gamma.\mview_{(w,q)}&\mbox{if
          $(w, a) \in \WR \times \RA$ 
         }\\
         \beta.\tview_t &\mbox{otherwise}
      \end{cases}} 
    }
    {\gamma, \beta\  \strans{a}_{t}\  \gamma[\tview_t \asgn \tview'], \beta[\tview_t \asgn \ctview'] }$
    \bigskip
    
  $ \inference[{\sc Write}] {
    a \in \{ wr(x,n), wr^{\sf R}(x,n)\} \qquad (w, q) \in \gamma.\OW(t, x) \setminus \gamma.\covered \qquad \fresh_\gamma(q,q') \\
    \writes' = \gamma.\writes \cup \{(a, q')\} \qquad
    \tview' = \gamma.\tview_t[x := (a, q')] \qquad \mview' = \tview' \cup \beta.\tview_t
    }
    {\gamma, \beta\  \strans{a}_{t}\  \gamma[\tview_t \asgn \tview', \mview_{(a,q')} \asgn \mview', \writes \asgn \writes'], \beta }$
    \bigskip
    
  $
    \inference[{\sc Update-RA}] {
    a = upd^{\sf
      RA}(x,m,n) 
    \qquad (w, q) \in \gamma.\OW(t, x) \setminus \gamma.\covered
    \qquad
    \wrval(w) = m 
    \qquad 
     \fresh_\gamma(q,q') \\ \writes' = \gamma.\writes \cup \{(a, q')\} \qquad
    \covered' = \gamma.\covered \cup \{(w, q)\} \qquad \mview' = \tview' \cup \ctview'
    \\
    \tview' = \ensuremath{
      \begin{cases}
        \gamma.\tview_t[x \asgn (a,
        q')]\otimes\gamma.\mview_{(w,q)}&\mbox{if $w \in
          \WR$ 
        }\\
        \gamma.\tview_t[x \asgn (a, q')]&\mbox{otherwise}
      \end{cases}}
    \\ 
    \ctview' = \ensuremath{
      \begin{cases}
        \beta.\tview_t \otimes\gamma.\mview_{(w,q)}&\mbox{if $w \in
          \WR$ 
        }\\
        \beta.\tview_t &\mbox{otherwise}
      \end{cases}}     
  }
    {\gamma, \beta\ \strans{a}_{t}\  \gamma\left[
      \begin{array}[c]{@{}l@{}}
        \tview_t \asgn \tview', \mview_{(a,q')} \asgn \mview', \\
        \writes \asgn \writes',
        \covered \asgn \covered'
      \end{array}\right], \beta[\tview_t\asgn \ctview']}$
  \bigskip

    $
    \inference[{\sc Update-Rel}] {
    a = upd^{\sf R}(x,m,n) 
    \qquad (w, q) \in \gamma.\OW(t, x) \setminus \gamma.\covered
    \qquad
    \wrval(w) = m 
    \\
     \fresh_\gamma(q,q') \qquad \writes' = \gamma.\writes \cup \{(a, q')\} \qquad
    \covered' = \gamma.\covered \cup \{(w, q)\} 
    \\
    \tview' = \gamma.\tview_t[x \asgn (a, q')] \qquad \mview' = \tview' \cup \beta.\tview_t
  }
    {\gamma, \beta\ \strans{a}_{t}\  \gamma\left[
      \begin{array}[c]{@{}l@{}}
        \tview_t \asgn \tview', \mview_{(a,q')} \asgn \mview', 
        \writes \asgn \writes',
        \covered \asgn \covered'
      \end{array}\right], \beta}$

  \caption{Transition relation for reads, writes and updates of the
  memory semantics}
  \label{fig:surrey-opsem}
\end{figure*}

\smallskip
\noindent{\bf Transition semantics.}
The transition relation of our semantics for global reads and writes
is given in \reffig{fig:surrey-opsem} and builds on an earlier
semantics that does not distinguish the state of the
context~\cite{ECOOP20}. Each transition
$\gamma, \beta \strans{a}_{t} \gamma', \beta'$ is labelled by an action
$a$ and thread $t$ and updates the target state $\gamma$ (the state of
component being executed) and the context
$\beta$. 

\smallskip
\noindent{\bf {\sc Read} transition by thread $t$.} Assume that $a$ is
either a relaxed or acquiring read to variable $x$, $w$ is a write to
$x$ that $t$ can observe (i.e., $(w, q) \in \gamma.\OW(t,x)$), and the
value read by $a$ is the value written by
$w$. 
Each read causes the viewfront of $t$ to be updated. For an
unsynchronised read, $\tview_t$ is simply updated to include the new
write. A synchronised read causes the executing thread's view of the
executing component and context to be updated. In particular, for each
variable $x$, the new view of $x$ will be the later (in timestamp
order) of either $\tview_t(x)$ or $\mview_w(x)$. To express this, we
use an operation that combines two views $V_1$ and $V_2$, by
constructing a new view from $V_1$ by taking the later view of each
variable: \begin{align*}
V_1 \otimes V_2  = \lambda x \in \dom(V_1).\ & 
    \textbf{if}~\tst(V_2(x)) \leq \tst(V_1(x))\  \textbf{then}\ V_1(x)\ \textbf{else}\  V_2(x)
  \end{align*}


\smallskip
\noindent{\bf {\sc Write} transition by thread $t$.} A write
transition must identify the write $(w,q)$ after which $a$
occurs. This $w$ must be observable and must {\em not} be covered ---
the second condition preserves the atomicity of read-modify-write (RMW)
updates. We must choose a fresh timestamp $q' \in \rat$ for $a$, which
for a C11 state $\gamma$ is formalised by $\fresh_\gamma(q, q') = q < q' \wedge \forall w' \in \gamma.\writes.\ q < \tst(w') \Rightarrow q' < \tst(w')$. 
That is, $q'$ is a new timestamp for variable $x$ and that
$(a,q')$ occurs immediately after $(w,q)$. The new write is added to
the set $\writes$. 

We update $\gamma.\tview_t$ to include the new
write, which ensures that $t$ can no longer observe any writes prior to
$(a, q')$. Moreover, we set the viewfront of $(a, q')$ to be the new
viewfront of $t$ in $\gamma$ together with the thread viewfront of the
environment state $\beta$. If some other thread synchronises with this
new write in some later transition, that thread's view will become at
least as recent as $t$'s view at this transition. Since $\mview$ keeps
track of the executing thread's view of both the component being
executed and its context, any synchronisation through this new write
will update views across components.

\smallskip
\noindent{\bf {\sc Update} (aka RMW) transition by thread $t$.} These
transitions are best understood as a combination of the read and write
transitions. As with a write transition, we must choose a valid fresh
timestamp $q'$, and the state component $\writes$ is updated in the
same way. State component $\mview$ includes information from the new
view of the executing thread $t$.  As discussed earlier, in {\sc
  Update} transitions it is necessary to record that the write that
the update interacts with is now covered, which is achieved by adding
that write to $\covered$. Finally, we must compute a new thread view,
which is similar to a {\sc Read} transition, except that the thread's
new view always includes the new write introduced by the
update. 

\begin{aside}[Generalisation to multiple components]
\label{aside:1}
The semantics can be extended to handle multiple environment
components, e.g., if a single client is interacting with multiple
libraries. Here, the context $\beta$ in \reffig{fig:surrey-opsem}
would be generalised to a set of states $S$. Release-acquire
synchronisation in the component $\gamma$, would lead to an updates in
the thread views for each of the context states in $S$. This would
lead to generalisation of the assertions in
\refsec{sec:example-verification} and the refinement rules in
\refsec{sec:cont-refin}.
\end{aside}


\section{Abstract object semantics}
\label{sec:abstr-object-semant}
The rules in \reffig{fig:surrey-opsem} provide a semantics for read,
write and update operations for component programs within an executing
context and can be used to model clients and libraries under RC11
RAR. These rules do not cover the behaviour of abstract objects, which
we now consider.


There have been many different proposals for specifying and verifying
concurrent objects
memory~\cite{DBLP:conf/pldi/Kokologiannakis19,DBLP:conf/popl/BattyDG13,DBLP:journals/pacmpl/EmmiE19,DBLP:conf/esop/KrishnaEEJ20,DBLP:journals/pacmpl/RaadDRLV19,ifm18,DongolJRA18},
since there are several different objectives that must be addressed.
These objectives are delicately balanced in
linearizability~\cite{HeWi90}, the most well-used consistency
condition for concurrent objects. Namely, linearizability ensures:
\begin{enumerate}
\item The abstract specification is explainable with respect to a
  \emph{sequential} specification.
\item Correctness is \emph{compositional}, i.e., any concrete
  execution of a system  comprising two linearizable objects is
  itself linearizable.
\item Correctness ensures \emph{contextual (aka observational)
    refinement}, i.e., the use of a linearizable implementation within
  a client program in place of its abstract specification does not
  induce any new behaviours in the client program.
\end{enumerate}

There is however an inherent cost to linearizability stemming from the
fact that the effect of each method call must take place \emph{before}
the method call returns. In the context of weak memory, this
restriction induces additional synchronisation that may not
necessarily be required for
correctness~\cite{DBLP:journals/cacm/SewellSONM10}. Therefore, over
the years, several types of relaxations to the above requirements have
been proposed
~\cite{DBLP:conf/pldi/Kokologiannakis19,DBLP:conf/popl/BattyDG13,DBLP:journals/pacmpl/EmmiE19,DBLP:conf/esop/KrishnaEEJ20,DBLP:journals/pacmpl/RaadDRLV19,ifm18,DongolJRA18}.




General data structures present many different design choices at the
abstract level~\cite{DBLP:journals/pacmpl/RaadDRLV19}, but discussing these now detracts from our main
contribution, i.e., the integration and verification of clients and
libraries in a weak memory model. 
Therefore, we restrict our attention to an abstract lock object, which
is sufficient to highlight the main ideas. Locks have a clear ordering
semantics (each new lock $\acquire$ and lock $\release$ operation must
have a larger timestamp than all other existing operations) and
synchronisation requirements (there must be a release-acquire
synchronisation from the lock \emph{\release} to the lock
\emph{\acquire}).

To enable proofs of contextual refinement (see
\refsec{sec:cont-refin}), we must ensure corresponding method calls
return the same value at the abstract and concrete levels. To this
end, we introduce a special variable $rval$ to each local state that
stores the value that each method call returns.
\begin{example}[Abstract lock]
  \label{ex:abs-lock}
  Consider the specification of a lock with methods 
  {\tt Acquire} and {\tt Release}. 
  Each method call of the lock is indexed by a subscript to uniquely
  identify the method call.  For the lock, the subscript is a counter
  indicating how many lock operations have been executed and is used
  in the example proof in \refsec{sec:example-verification}.
  \[ 
    \inference[{\sc Acq}] {a = l.\acquire_n \quad ls' = ls[rval := \True] 
    } 
    {({\tt l.Acquire()}, ls) \trans a (\True, ls')}
  \quad \ \ \inference[{\sc Rel}]
    {a = l.\release_n \quad ls' = ls[rval := \bot]} 
    {({\tt l.Release()}, ls) \trans a (\bot, ls')}
  \]

  \noindent
  Locks, by default are synchronising, i.e., in the memory
  semantics, a (successful) \emph{\acquire} requires the operation to
  synchronise with the most recent lock \emph{\release} (in a manner
  consistent with release-acquire semantics), so that any writes that
  are happens-before ordered before the \emph{\release} are visible to
  the thread that acquires the lock. The initial state of an abstract
  lock $l$, $\gamma_\Init$, is given by: 
  \begin{align*}
    \gamma_\Init.\writes & = \{(l.init_0,0)\}
    & 
    \alpha_\Init.\tview_t(l) & =  (l.init_0,0)
    & 
    \alpha_\Init.\covered & =  \emptyset
  \end{align*}  
  We also obtain the rules below, where  $\alpha$ is the state of
  the lock and $\gamma$ is the state of the client.

To record the thread that currently owns the lock, we derive a new
  action, $b$, from the action $a$ of the program
  semantics. Action $(w, q)$ represents the method that is observed by
  the $\acquire$ method, which must be an operation in
  $\alpha.\writes$ such that $q$ has the maximum timestamp for $l$
  (i.e., $q = \maxTS(\alpha.\writes_{|l})$). The new timestamp $q'$ must be larger
  than $q$. We create a new component state $\alpha'$ from $\alpha$ by (1) inserting $(b, q')$ into $\alpha.\writes$; (2) updating $\tview_t$ to $\tview'$, where $\tview'$
    synchronises with the previous thread view in $\alpha$ to include
    information from the modification view of $(w, q)$, and updates $t$'s view of $l$ to include the new operation $(b, q')$;  (3)  updating the contextual thread view for $t$ to $\ctview'$,
    where $\ctview'$ synchronises with the previous thread view in the
    context state $\gamma$ to include information from the modification
    view of $(w, q)$; and (4) updating the modification view for the new operation $(b, q')$
    to $\mview'$, where $\mview'$ contains the view of $t$.
  Finally, the context state $\gamma'$ updates the thread view of $t$ to
  $\ctview'$ since synchronisation with a release may cause the view
  to be updated.
  
  A lock release, simply introduces a new operation with a maximal
  timestamp, provided that the thread executing the release currently holds
  the lock.
\end{example}

  \begin{figure}[t]
    \centering \small
    $    \inference[{\sc Acquire}] {
      a = l.\acquire_n  \qquad b = l.\acquire_n(t) \qquad (w, q) \in \alpha.\writes \qquad  w \in \{l.init_0, l.\release_{n-1}\} \\
      q = \maxTS(\alpha.\writes_{|l}) \qquad q < q' \\ \writes' = \alpha.\writes
      \cup \{(b, q')\} 
      \qquad \mview' = \tview' \cup \ctview' \qquad \covered' = \sigma.\covered \cup \{(w,q)\} \\
      \tview' = \alpha.\tview_t[l \asgn (b,
      q')]\otimes\alpha.\mview_{(w,q)} \quad \ctview' = \gamma.\tview_t
      \otimes\alpha.\mview_{(w,q)} \\
      } {\alpha, \gamma\ \strans{a}_{t}\
      \alpha\left[
        \begin{array}[c]{@{}l@{}}
          \writes := \writes', \tview_t := \tview', 
          \\
          \mview_{(b, q')} := \mview', \covered := \covered'
        \end{array}\right]
      , \gamma[\tview_t \asgn \ctview']}
    $

    \bigskip
    $
    \inference[{\sc Release}] {
        a = l.\release_n \quad w = l.\acquire_{n-1}(t) \quad 
        (w, q) \in \alpha.\writes \quad
        q = \maxTS(\alpha.\writes_{|l}) \quad q < q'  \\ 
        \writes' = \alpha.\writes \cup \{(a, q')\} \qquad
        \tview' = \alpha.\tview_t[x \asgn (a, q')]
        \\ 
        \mview' = \tview' \cup \gamma.\tview_t
      } 
      {\alpha, \gamma\  \strans{a}_{t}\  \alpha\left[
          \begin{array}[c]{@{}l@{}}
            \writes := \writes', \tview_t := \tview', 
            \mview_{(a, q')} := \mview'
          \end{array}\right]
        , \gamma}
      $

\caption{Operational semantics for lock acquire and release }
    \label{fig:acq-rel-sem}
  \end{figure}







\newcommand{\lockver}{{\it lv}}
\section{Client-library verification}
\label{sec:example-verification}

Having formalised the semantics of clients and libraries in a weak
memory setting, we now work towards verification of (client) programs
that use such libraries.  We develop an assertion language for
specifying client-library states in \refsec{sec:assertion-language}, a
Hoare logic for reasoning about client-library programs in
\refsec{sec:hoare-logic-c11} and an example verification of a client
program using a lock library in \refsec{sec:example-client-lbjec}.


\subsection{Assertion language}
\label{sec:assertion-language}
In our proof, we use \emph{observability assertions}, which describe
conditions for a thread to observe a specific value for a given 
variable. Unlike earlier works, our operational semantics covers
clients and their libraries, and hence operates over pairs of
states. 

 \smallskip
\noindent
{\bf Possible observation}, denoted $\langle x = u\rangle_t$, means
that thread $t$ \emph{may} observe value $u$ for
$x$~\cite{ECOOP20}. We extend this concept to cope with abstract
method calls as follows. In particular, for an object $o$ and method
$m$, we use $\langle o.m \rangle_t$ to denote that thread $t$ can observe
$o.m$.
  \begin{align*}
    \langle x = n \rangle_t (\sigma)  & \ \  \equiv \ \ \exists w\in
                              \sigma.\OW(t,x).\ \wrval(w) = n\\
    \langle o.m \rangle_t (\sigma) & \ \  \equiv \ \ \exists q.\ (o.m, q) \in
                            \sigma.\writes \wedge q \geq \sigma.\tview_t(o)
  \end{align*}
  
  \noindent To distinguish possible observation in clients and libraries, we introduce the following notation,
  where $\gamma$ and $\beta$ are the client and library states,
  respectively, and $p$ is either a valuation (i.e., $x = n$) or an
  abstract method call (i.e., $o.m$):
  \begin{align*}
    \langle p \rangle_t^C (\gamma, \beta) & \ \ \equiv \ \ \langle p \rangle_t (\gamma)  &
                                                                                               \langle p \rangle_t^L (\gamma, \beta) & \ \ \equiv \ \ \langle p \rangle_t(\beta)
  \end{align*}
  
\smallskip
\noindent {\bf Definite observation}, denoted $[x = u]_t$, means that
thread $t$ can only see the last write to $x$, and that this write has
written value $u$. Recall that each write is a pair containing a write
event and a timestamp. Let $W$ be a set of writes with a unique
timestamp for each $x$, i.e., $(W_{|x})^{-1}$ is a function.  We
define the \emph{last write} to $x$ in a set of writes $W$ as:
  \begin{align*}
    \last(W, x) \ \  \equiv \ \  & (W_{|x})^{-1}(\maxTS(W_{|x})) 
  \end{align*}
  We define the definite observation of a view function, $view$ with
  respect to a set of writes as follows:
  \begin{align*}
    \begin{array}[t]{@{}r@{}l@{}}
      \dview (view, W, x) = n 
      \ \   \equiv\ \  &  view(x) = \last(W, x) \wedge 
    \wrval(\last(W, x)) = n
    \end{array}
  \end{align*}
  The first conjunct ensures that the viewfront of $view$ for $x$ is the
  last write to $x$ in $W$, and the second conjunct ensures that the
  value written by the last write to $x$ in $W$ is $n$.  For a variable
  $x$, thread $t$ and value $n$, we define:
  \begin{align*}
    [x = n]_t (\sigma) \ \ & \equiv \ \ \dview(\sigma.\tview_t, \sigma.\writes \cap \W, x) =n
  \end{align*}
  The extension of definite observation assertions to abstract method
  calls is straightforward to define. Namely we have: 
  \begin{align*}
    [o.m]_t (\sigma) \ \ & \equiv \ \ 
                  \begin{array}[t]{@{}l@{}}
                    \sigma.\tview_t(o) =  \maxTS(\sigma.\writes_{|o}) \wedge {} 
                    (o.m, \maxTS(\sigma.\writes_{|o})) \in \sigma.\writes
                  \end{array}
  \end{align*}
  As with possible observations, we lift definite observation
  predicates to state spaces featuring clients and libraries: 
  \begin{align*}
    [p]^C_t (\gamma, \beta) & \ \ \equiv \ \ [p]_t(\gamma) & 
                                                                 [p]^L_t (\gamma, \beta) & \ \ \equiv \ \ [p]_t(\beta)
  \end{align*}

\smallskip
\noindent {\bf Conditional observation}, denoted
$\CObs{x}{u}{t}{y}{v}$, means that if thread $t$ synchronises with a
write to variable $x$ with value $u$, it \emph{must} subsequently
observe value $v$ for $y$.  For variables $x$ and $y$, thread $t$ and
values $u$ and $v$, we define:
\begin{align*}
                         \begin{array}[t]{@{}r@{~}l@{}}
                           & \CObs{x}{u}{t}{y}{v} (\sigma) \\
                           \ \ \equiv \ \ & 
                           \forall w \in \sigma.\OW(t,x) .\
                           \wrval(w) = u \imp  \act(w) \in \WR \wedge \dview(\sigma.\mview_w, \sigma.\writes, y) = v 
                         \end{array}
\end{align*}
This is a key
assertion used in message passing proofs
\cite{ECOOP20,DBLP:conf/ecoop/KaiserDDLV17} since it guarantees an
observation property on a variable, $y$, via a synchronising read of
another variable, $x$.


As with possible and definite assertions, conditional assertions can be
generalised to objects and extended to pairs of states describing a
client and its library. However, unlike possible and definition
observations assertions, conditional observation enables one to
describe view synchronisation across different states. For example,
consider the following, which enables conditional observation of an
abstract method to establish a definite observation assertion for the
thread view of the client.  We assume a set $Sync \subseteq \Act$ that
identifies a set of synchronising abstract actions.
\begin{align*}
  \begin{array}[t]{@{}r@{~}l@{}}
    \langle o.m \rangle^L[y = v]_t^C (\gamma, \beta)  \ \ \equiv \ \  &
    \forall q.\ (o.m, q) \in \beta.\writes \wedge q \geq \beta.\tview_t(o) \imp \\
&       {} \qquad\qquad o.m \in Sync \wedge \dview(\beta.\mview_{(o.m, q)}, \gamma.\writes, y) = v
    \end{array}
\end{align*}
It is possible to define other variations, e.g., conditional observation
synchronisation from clients to libraries, but we leave out the
details of these since they are straightforward to construct. 

\smallskip
\noindent {\bf Covered operations}, denoted $\cvd{o.m}{}$, holds iff
all but the last operation $m$ of the object $o$ is covered. 
Recall from
the {\sc Acquire} rule that
a new acquire operation causes the immediately prior (release)
operation $l.release_{n-1}$ to be covered so that no later acquire can
be inserted between $l.release_{n-1}$ and the new acquire. 
To reason about this phenomenon over states, we use:
\begin{align*}
  \cvd{o.m}{} (\sigma) &\ \  \equiv\ \
                         \begin{array}[t]{@{}l@{}}
                           \forall (w, q) \in \sigma.\writes_{|o} \setminus \sigma.\covered.\   w = o.m \wedge q = \maxTS(\sigma.\writes_{|o})
                         \end{array}
\end{align*}
where $\sigma.\writes_{|o}$ is the set of operations over object
$o$.

\smallskip
\noindent {\bf Hidden value}, denoted $\cvv{o.m}{}$, states that
the operation $o.m$ exists, but all of these are hidden from
interaction. 
In proofs, such assertions limit the values that
can be returned. 
\begin{align*}
  \cvv{o.m}{}(\sigma)
  &\ \  \equiv\ \  
    \begin{array}[t]{@{}l@{}}
      (\exists q.\ (o.m, q) \in \sigma.\writes) \wedge {}  (\forall q.\ (o.m, q) \in \sigma.\writes  \imp (o.m, q) \in \sigma.\covered)
    \end{array} 
\end{align*}
Both covered and hidden-value assertions can be lifted to pairs of
states and can be used to reason about standard writes, as opposed to
method calls (details omitted). For instance, we can define
$\cvd{x}{C}(\gamma, \beta) \ \equiv\ \cvd{x}{}(\gamma)$ and
$\cvd{x}{L}(\gamma, \beta) \ \equiv\ \cvd{x}{}(\beta)$.

\subsection{Hoare Logic and Owicki-Gries for C11 and Abstract Objects}
\label{sec:hoare-logic-c11}
Since we have an operational semantics, the assertions in \refsec{sec:assertion-language} can be
integrated into standard Hoare-style proof calculus in a
straightforward
manner~\cite{ECOOP20,DBLP:journals/corr/abs-2004-02983}. The only
differences are the state model (which is a weak memory state, as
opposed to mappings from variables to values) and the atomic
components (which may include reads of global variables, and, in this
paper, abstract method calls).

Following \cite{ECOOP20,DBLP:journals/corr/abs-2004-02983}, we let
$\Sigma_C$ and $\Sigma_{L}$ to be the set of all possible global
state configurations of the client and library, respectively and let
$\Sigma_{C11} = (\LVar \to \Val) \times \Sigma_C \times \Sigma_{L}$
be the set of all possible client-library C11 states. Predicates over
$\Sigma_{C11}$ are therefore of
type $\Sigma_{C11} \to \mathbb{B}$.  This leads to the following
definition of a Hoare triple, which we note is the same as the
standard definition --- the only difference is that the state
component is of type $\Sigma_{C11}$.
\begin{definition}
  \label{def:soundn-class-rules-1}Suppose 
  $p, q \in \Sigma_{C11} \to \mathbb{B}$, $P \in {\it Prog}$ and
  $\textsf{\textbf{E}} = \lambda t : \Tid.\ \kwskip$.  The semantics
  of a Hoare triple under partial correctness is given by: \smallskip
  
  $\begin{array}{r@{~}l}
    \{p\} \Init \{q\} & =  q(\Gamma_\Init) 
    \\[2pt]
        \{p\}
\Init ; P \{q\} & =  \exists r.\,\{p\} 
                      \Init \{r\} \wedge \{r\} P \{q\}

     \\[2pt]
                   \{p\}P \{q\} & = \begin{array}[t]{@{}l@{}}
                                      \forall \lstate,\gamma, \beta ,\lstate',\gamma', \beta'.\ \\
                                      \qquad p(\lstate,
                   \gamma, \beta) \wedge     ((P, \lstate, \gamma, \beta) \ltsArrow{}^* (\textsf{\textbf{E}},
                   \lstate', \gamma', \beta')) \	\imp \	 q(\lstate',\gamma', \beta')
                   \end{array}
    \\
  \end{array}$
\end{definition}

This definition (in the context of RC11
\cite{DBLP:conf/pldi/LahavVKHD17}) allows all standard Hoare logic
rules for compound statements to be reused \cite{ECOOP20}.  Due to
concurrency, following Owicki and Gries, one must prove \emph{local
  correctness} and \emph{interference freedom} (or
stability)~\cite{DBLP:journals/acta/OwickiG76,ECOOP20,DBLP:journals/corr/abs-2004-02983,DBLP:conf/icalp/LahavV15}. This
is also defined in the standard manner.
Namely, a statement $R \in \AComm$ with precondition $pre(R)$ (in the
standard proof outline) does {\em not interfere} with an assertion $p$
iff $\{ p \wedge pre(R)\} \ R \ \{ p\}.$ Proof outlines of concurrent
programs are {\em interference free} if no statement in one thread
interferes with an assertion in another thread.

The only additional properties that one must define are on the
interaction between atomic commands and predicates over assertions
defined in \refsec{sec:assertion-language}. A collection of rules for
reads, writes and updates have been given in prior
work~\cite{DBLP:journals/corr/abs-2004-02983,ECOOP20}. Here, we
present rules for method calls of the abstract lock object defined in
\refex{ex:abs-lock}.

In proofs, it is often necessary to reason about particular versions
of the lock (i.e., the lock counter). Therefore, we use ${\tt l.Acquire(\mbox{$v$})}$
and ${\tt l.Release(\mbox{$v$})}$ to denote the transitions that set the lock version to
$v$. Also note that in our example proof, it is clear from context
whether an assertion refers to the client or library state, and hence,
for clarity, we drop the superscripts $C$ and $L$ as used in
\refsec{sec:assertion-language}.

The lemma below has been verified in Isabelle/HOL.

\begin{lemma} Each of the following holds,  
  where the statements are
  decorated with the  identifier of the executing thread, assuming ${\tt m} \in \{{\tt Acquire}, {\tt Release}\}$ and $t \neq t'$
  \begin{small}
  \begin{align} 
  & \{\cvv{l.\release_u}{}\}\ {\tt
    l.Acquire(\mbox{$v$})_t} \ \{ v > u + 1 \} 
    \\
   & \{\cvv{l.\release_u}{}\}\ {\tt l.m(\mbox{$v$})_t}\
    \{\cvv{l.\release_u}{}\} 
    \\
   & \{[l.\release_u]_t\}
    \ {\tt
    l.Acquire(\mbox{$v$})_t}\ \{[l.\acquire_{u+1}]_t\}\\
   & \{[x = u]_t\}\ {\tt l.m(\mbox{$v$})_{t'}}\
    \{[x = u]_t\} \\
   & \{
       \langle l.\release_u \rangle [x = n]_t \}\
     {\tt l.Acquire(\mbox{$v$})_t}\ \{v = u + 1 \imp [x = n]_t\} 
    \\
   &  \left\{
     \begin{array}[c]{@{}l@{}}
       \neg \langle l.\release_u \rangle_{t'} 
        {}\wedge{} {[}x = v]_t
     \end{array}
\right\}\ {\tt l.Release(\mbox{$u$})_{t}}\
     \{\langle \release_u \rangle [x = v]_{t'}\} 
     \label{co-intro}
  \end{align}
\end{small}


\end{lemma}
  








\vspace{-2em}

\subsection{Example Client-Library Verification}
\label{sec:example-client-lbjec}
To demonstrate use of our logic in verification, consider the simple
program in \reffig{fig:lockmp-proof}, which comprises a lock object
$l$ and shared client variables $d_1$ and $d_2$ (both initially~$0$).
Thread~1 writes 5 to both $d_1$ and $d_2$ after acquiring the lock
while thread~2 reads $d_1$ and $d_2$ (into local registers $r_1$ and
$r_2$) also after acquiring the lock.

Under SC, it is a standard exercise to show that
the program terminates with $r_1 = r_2$ and $r_i = 0$ or $r_i =
5$. 
We show that the lock specification in
\refsec{sec:abstr-object-semant} together with the assertion language
from \refsec{sec:assertion-language} and Owicki-Gries logic from
\refsec{sec:hoare-logic-c11} is sufficient to prove this property. In
particular, the specification guarantees {\em adequate synchronisation} so
that if the Thread~2's lock acquire sees the lock release in Thread 1,
it also sees the writes to $d_1$ and $d_2$. The proof 
relies on two distinct types of properties:
\begin{itemize}[leftmargin=*]
\item \emph{Mutual exclusion}: As in SC, no
  two threads should execute their critical sections at the same
  time.
\item \emph{Write visibility}: If thread 1 enters its critical section
  first, its writes to both $d_1$ and $d_2$ must be visible to thread
  2 after thread 2 acquires the lock. Note that this property is not
  necessarily guaranteed in a weak memory setting since all reads/writes
  to $d_1$ and $d_2$ in \reffig{fig:lockmp-proof} are relaxed.
\end{itemize}

\begin{figure*}[t]
  \centering
\fbox{
\begin{minipage}[b]{0.7\textwidth}
    \begin{center}  \small
  {\bf Init: } $d_1:=0;$ $d_2:=0;$ ${\tt l.init()};$  \qquad \qquad \qquad \qquad\\ 
   {\color{green!40!black} $\{Inv \wedge [d_1 = 0]_1 \wedge [d_2 = 0]_1 \wedge [d_1 = 0]_2 \wedge [d_2 = 0]_2\}$} \\ 
  $\begin{array}{l@{\quad}||@{\quad }l}
     \text{\bf Thread } 1
     & \text{\bf Thread } 2\\
     
     \begin{array}[t]{@{}l@{}}
       1: \{Inv \wedge \mathbf{P_1}\}\ {\bf if}\ {\tt l.Acquire()} \\ 
       2: \{Inv \wedge \mathbf{P_2}\}\ \quad d_1 := 5 ; \\
       3: \{Inv \wedge \mathbf{P_3}\}\  \quad d_2 := 5 ; \\
       4: \{Inv \wedge \mathbf{P_4}\}\ \quad {\tt l.Release()} 
     \end{array}
     & 
     \begin{array}[t]{@{}l@{}}
       1: \{Inv \wedge \mathbf{Q_1}\}\ {\bf if}\ {\tt l.Acquire(\mbox{$rl$})} \\ 
       2: \{Inv \wedge \mathbf{Q_2}\}\ \quad r_1 \gets d_1 ; \\
       3: \{Inv \wedge \mathbf{Q_3}\}\ \quad r_2 \gets d_2 ; \\
       4: \{Inv \wedge \mathbf{Q_4}\}\ \quad{\tt l.Release()} 
     \end{array}
     \end{array}$

   {\color{green!40!black} $5: \{(r_1=0 \wedge r_2=0) \vee (r_1=5 \wedge r_2=5)\}$}  \qquad \qquad  \qquad      
 \end{center}
\end{minipage}}
\medskip

\raggedright \small
where assuming $P_{po}  =  (pc_2 = 1 \imp  \neg \langle l.\release_2 \rangle_2)\wedge \cvv{l.\init_0}{}$, we have  

\centering
$
\begin{array}[t]{@{}l@{\quad}l@{}}
 \begin{array}[t]{@{}r@{}l@{}}
   P_1 =  &   [d_1 = 0]_1
     \wedge [d_2 = 0]_1 
           \wedge  \\
  & (pc_2 = 1 \imp [l.init_0]_{1} \wedge [l.init_0]_{2})\\
  &  \wedge  (pc_2 \in \{2,3,4\} \imp \cvd{l.\acquire_1}{}) \\
  P_2 =   & [d_1 = 0]_1 \wedge [d_2 = 0]_1
   \wedge P_{po}
\\
  P_3 =  & [d_1 = 5]_1 \wedge [d_2 = 0]_1
   \wedge P_{po}
        \\
  P_4  =  & [d_1 = 5]_1 \wedge [d_2 = 5]_1
   \wedge
   P_{po}
   \\
   \end{array}
        & 
          \begin{array}[t]{@{}r@{}l@{}}
            & Q_1' = pc_1 = 5 \wedge \langle l.\release_2 \rangle [d_1 = 5]_2 {} \wedge \langle l.\release_2\rangle [d_2 = 5]_2 \\
     & Q_1  =  \begin{array}[t]{@{}l@{}}
   \left(
   \begin{array}[c]{@{}l@{}l@{}}
     pc_1 \notin \{2,3,4\} \imp \\
     \qquad ([l.init_0]_2 \wedge [d_1 = 0]_2 \wedge  {[}d_2 = 0]_2)   {} \lor  Q_1' {} \\
   \end{array}\right)
   \\
   {} \wedge (pc_1 = 1 \imp [l.init_0]_1) 
   \wedge (pc_1 = 5 \imp \cvv{l.\init_0}{})
 \end{array} 
\\
 & Q_2  =  (rl = 1 \imp [d_1 = 0]_2 \wedge [d_2 = 0]_2)
    \\
 & \qquad\ \  {} \wedge (rl = 3 \imp [d_1 = 5]_2 \wedge [d_2 = 5]_2)   
\\
 & Q_3  =
   \begin{array}[t]{@{}l@{}}
     (rl = 1 \imp r_1 = 0 \wedge [d_2 = 0]_2)  \\
     {} \wedge (rl = 3 \imp r_1 = 5 \wedge [d_2 = 5]_2)
   \end{array}
        \\
 & Q_4  =
   \begin{array}[t]{@{}l@{}}
(rl = 1 \imp r_1 = 0 \wedge r_2 = 0) \\
 {} \wedge (rl = 3 \imp r_1 = 5 \wedge r_2 = 5)
   \end{array}

        \end{array}
\end{array}
$
 \caption{Proof outline for lock-synchronisation}
\label{fig:lockmp-proof}
\end{figure*}

Our proof is supported by the following global invariant:
\begin{align*}
  Inv \ \ \equiv{}\ \  & \neg (pc_1 \in \{2,3,4\} \wedge pc_2 \in \{2,3,4\}) \wedge (rl \in \{1, 3\})
\end{align*}
The first conjunct establishes mutual exclusion, while the second
ensures that the lock version written by the acquire in thread 2 is
either 1 or 3, depending on which thread enters its critical section
first.
The main purpose of the
definite and possible observation assertions is to establish 
$Q_1'$ (which appears in $Q_1$) using rule \refeq{co-intro}.
This predicate helps establish $[d_1 = 5]_2$ and
$[d_2 = 5]_2$ in thread 2 whenever thread 2 acquires the lock after
thread~1. The most critical of these assertions is $Q_1$, which states that if
thread 1 is not executing its critical section then we either have
\begin{itemize}[leftmargin=*]
\item $[l.init_0]_2 \wedge [d_1 = 0]_2 \wedge [d_2 = 0]_2$, i.e.,
  thread 2 can definitely see the lock initialisation and definitely
  observe both $d_1$ and $d_2$ to have value $0$, or
\item 
  $Q_1'$ holds,
  i.e., thread 1 has
  released the lock and has established a state whereby if thread 2
  acquires the lock, it will be able to establish the definite value
  assertions $[d_1 = 5]_2$ and $[d_2 = 5]_2$.
\end{itemize}
Note that $Q_1$ also includes a conjunct
$pc_1 = 5 \imp \cvv{l.\init_0}{}$, which ensures that if thread 2 enters
its critical section after thread 1 has terminated, then it does so
because it sees $l.\release_2$ (as opposed to $l.\init_0$). This means
that we can establish $rl = 1 \imp [d_1 = 0]_2 \wedge [d_2 = 0]_2$
(i.e., thread 2 has acquired the lock first) and
$rl = 3 \imp [d_1 = 5]_2 \wedge [d_2 = 5]_2$ (i.e., thread 2 has
acquired the lock second) in $Q_2$. Using these definite value
assertions, we can easily establish that the particular values that
are loaded into registers $r_1$ and $r_2$. The lemma has been verified
in Isabelle/HOL.

\begin{lemma}
  \label{thm:lockmp-proof}
  The proof outline in \reffig{fig:lockmp-proof} is valid. 
\end{lemma}

\newcommand{\Tr}{{\it Tr_{SF}}}
\newcommand{\ct}{{\it ct}}
\newcommand{\at}{{\it at}}
\newcommand{\refby}{\sqsubseteq}
\newcommand{\crefby}{\mathrel{\widehat{\sqsubseteq}}}
\newcommand{\glo}{{\it glo}}
 \newcommand{\state}{{\it state}}

\section{Contextual Refinement}
\label{sec:cont-refin}

We now describe what it
means to \emph{implement} a specification so that any client
property that was preserved by the specification is not invalidated by the
implementation. 
We define
and prove contextual refinement directly, i.e., without appealing to
external correctness conditions over libraries, cf.
linearizability~\cite{HeWi90,ifm18,GotsmanY11,DongolJRA18,DBLP:journals/tcs/FilipovicORY10}. 
We present our refinement theory in \refsec{sec:refin-simul-weak},
including a proof technique based on forward simulation.
In Sections
\ref{sec:sequence-lock} and \ref{sec:ticket-lock}, we provide two
examples of contextual refinement of the abstract lock object
introduced in \refsec{sec:abstr-object-semant}. 

\subsection{Refinement and Simulation for Weak Memory}
\label{sec:refin-simul-weak}

Since we have an operational semantics with an interleaving semantics
over weak memory states, the development of our refinement theory
closely follows the standard approach under
SC~\cite{DBLP:books/cu/RoeverE1998}.
Suppose $P$ is a program with initialisation $\Init$. An
\emph{execution} of $P$ is defined by a possibly infinite sequence
$\Pi_0\, \Pi_1\, \Pi_2\,\dots$ such that
\begin{enumerate}[leftmargin=*]
\item each $\Pi_i$ is a 4-tuple $(P_i, ls_i, \gamma_i, \beta_i)$
  comprising a program, local state, global client state and global library state, and
\item
  $(ls_0, \gamma_0, \sigma_0) = (\ls_\Init, \gamma_\Init,
  \sigma_\Init)$, and
\item for each $i$, we have $\Pi_i \Longrightarrow \Pi_{i+1}$ as
  defined in \refsec{sec:program-semantics}.
\end{enumerate}
A \emph{client trace} corresponding to an execution
$\Pi_0\,\Pi_1\,\Pi_2\dots$ is a sequence $\ct \in \Sigma_C^*$
such that $\ct_i = (\pi_{2}(\Pi_i)_{|C}, \pi_{3}(\Pi_i))$, where
$\pi_n$ is a projection function that extracts the $n$th component of
a given tuple and $ls_{|C}$ restricts the given local state $ls$ to
the variables in $\LVar_C$.  Thus each $\ct_i$ is the global client
state component of $\Pi_i$.

After a projection, the concrete implementation may contain (finite or
infinite) stuttering~\cite{DBLP:books/cu/RoeverE1998}, i.e.,
consecutive states in which the client state is unchanged. We let
${\it rem\_stut}(\ct)$ be the function that removes all stuttering
from the trace $\ct$, i.e., each consecutively repeating state is
replaced by a single instance of that state. We let $\Tr(P)$ denote
the set of \emph{stutter-free traces} of a program $P$, i.e., the
stutter-free traces generated from the set of all executions of $P$.

Below we refer to the client that uses the abstract object as the
\emph{abstract client} and the client that uses the object's
implementation as the \emph{concrete client}. The notion of contextual
refinement that we develop ensures that a client is not able to distinguish 
the use of a concrete implementation in place of an abstract
specification. In other words, each thread of the concrete client
should only be able to observe the writes (and updates) in the client
state (i.e., $\gamma$ component) that the thread could already observe
in a corresponding of the client state of the abstract client.
First we define trace refinement for weak memory states. 
\begin{definition}[State and Trace Refinement]
  \label{def:cont-refin-1}
  We say a concrete state $\gamma_C$ is a \emph{refinement} of an
  abstract state $\gamma_A$, denoted
  $(ls_A, \gamma_A) \sqsubseteq (ls_C, \gamma_C)$ iff $ls_A = ls_C$,
  $\gamma_A.\covered = \gamma_C.\covered$ and for all threads $t$ and
  $x \in \GVar$, we have
  $\gamma_C.\OW(t, x) \subseteq \gamma_A.\OW(t, x)$. 
  We say a concrete
  trace $\ct$ is a \emph{refinement} of an abstract trace $\at$,
  denoted $\at \refby \ct$, iff $\ct_i \sqsubseteq \at_i$ for all $i$.
\end{definition}
This now leads to a natural trace-baseed definition of contextual refinement. 
\begin{definition}[Program Refinement]
  \label{def:prog-ref}
  A concrete program $P_C$ is a \emph{refinement} of an abstract
  program $P_A$, denoted $P_A \refby P_C$, iff for any (stutter-free)
  trace $\ct \in \Tr(P_C)$ there exists a (stutter-free) trace
  $\at \in \Tr(P_A)$ such that $\at \refby \ct$.
\end{definition}
Finally, we obtain a notion of contextual refinement for abstract
objects. Suppose $P$ is a program with holes. We let $P[O]$ be the
program in which the holes are filled with the operations from object
$O$. Note that $O$ may be an abstract object, in which case execution
of each method call follows the abstract object semantics
(\refsec{sec:abstr-object-semant}), or a concrete implementation, in
which case execution of each method call follows the semantics of
reads, writes and updates (\refsec{sec:program-semantics}).
\begin{definition}[Contextual refinement]
  \label{def:cref}
  We say a concrete object $CO$ is a \emph{contextual refinement} of
  an abstract object $AO$ 
   iff for any client
  program $C$, we have $C[AO] \refby C[CO]$.
\end{definition}

To verify contextual refinement, we use a notion of \emph{simulation},
which once again is a standard technique from the literature. The
difference in a weak memory setting is the fact that the refinement
rules must relate more complex configurations, i.e., tuples of the
form $(P, {\it lst}, \gamma, \alpha)$. 

The simulation relation, $R$, relates triples
$(als, \gamma_A, \alpha)$, comprising an abstract local state $als$,
client state $\gamma_A$ and library state $\alpha$, with triples
$(cls, \gamma_C, \beta)$ comprising a concrete local state $cls$, a
client state $\gamma_C$ and concrete library state $\beta$. The
simulation condition must ultimately ensure
$(als_{|C}, \gamma_A) \refby (cls_{|C}, \gamma_C)$ at each step as
defined in \refdef{def:cont-refin-1}. However, since client
synchronisation can affect the library state, a generic forward
simulation rule is non-trivial to define since it requires one to
describe how clients steps affect the simulation relation. We
therefore present a simpler use case for libraries that are used by
clients that do not perform any synchronisation outside the library
itself (e.g., the client in \reffig{fig:lockmp-proof}). If
$\Pi = (P, lst, \gamma, \alpha)$, we let
$\state(\Pi) = (lst, \gamma, \alpha)$ be the state corresponding to
$\Pi$.


\begin{definition}[Forward simulation for synchronisation-free
  clients] \label{def:fsim} For an abstract object $AO$ and a concrete object $CO$ and
  a client $C$ that only synchronises through $AO$ (and $CO$),
  $C[AO] \refby C[CO]$ holds if there exists a relation $R$ such that
  \begin{enumerate}[leftmargin=*]
  \item $R((als, \gamma_A, \alpha), (cls, \gamma_C, \beta)) \imp $

    $
    \begin{array}[c]{l}
      als_{|C} = cls_{|C} 
      \wedge {} \\
      \forall t, x.\
      \begin{array}[t]{@{}l@{}}
\gamma_C.\OW(t, x) \subseteq \gamma_A.\OW(t, x) \wedge  
      als(t)(rval) = cls(t)(rval)
      \end{array}
    \end{array}
$
  \hfill   (client observation)
  \item
    $R(\state(\Omega_\Init), \state(\Pi_\Init))$ \hfill (initialisation)
  \item For any concrete configurations $\Pi$, $\Pi'$ and abstract
    configuration $\Omega$, if $\Pi \Longrightarrow \Pi'$ via a step
    corresponding to $CO$, and $R(\state(\Omega), \state(\Pi))$, then either
    \begin{itemize}
    \item $R(\state(\Omega), \state(\Pi))$, or \hfill (stuttering step)
    \item there exists an abstract configuration $\Omega'$ such that
      $\Omega \Longrightarrow \Omega'$ and $R(\state(\Omega'),
      \state(\Pi'))$.
      
      \hfill (non-stuttering step)
    \end{itemize}
  \end{enumerate}
\end{definition}

\begin{theorem} If $R$ is a forward simulation between $AO$ and $CO$,
  then for any client that only synchronises through $AO$ (and $CO$)
  we have $C[AO] \refby C[CO]$.
\end{theorem}

\subsection{Sequence Lock}
\label{sec:sequence-lock}



The first refinement example (\reffig{seqlock_imp}) is a sequence
lock, which operates over a single shared variable ($glb$). The
${\bf Acquire}$ operation returns true if, and only if, the $\kwcas$
on line~2 is successful. Therefore, in order to prove the refinement,
we will need to prove that whenever the $\kwcas$ operation is
successful, the abstract object can also successfully acquire the lock
maintaining the simulation relation.  Also, the read on line 1 and the
unsuccessful $\kwcas$ are stuttering steps and we need to show that
when those steps are taken the abstract state remains unchanged and
the new concrete state preserves the simulation relation. The
${\bf Release}$ operation contains only one releasing write on
variable $glb$, which is considered to be a refining step. It is
straightforward to show that this operation refines the abstract
object release operation. The following proposition has been verified
in Isabelle/HOL.

\begin{figure}[t]
    \begin{minipage}[b]{0.48\textwidth}
      \textbf{Init:} \ \ glb = 0
      \\[6pt]
      \begin{minipage}[t]{\textwidth}
\begin{minipage}[b]{0.80\textwidth}
  \textbf{Acquire()}: 
        \begin{algorithmic}[1] \small
          \State \textbf{do} \quad \textbf{do}
            r $\leftarrow^{\sf A}$ glb 
          \textbf{until} even(r) ;
          \State \qquad\ \   loc $ \gets$ $\kwcas^{\sf RA}$(glb, r, r+1)
          \State \textbf{until} loc
        \end{algorithmic}
      \end{minipage}
      \end{minipage}
      \\[6pt]     
      \begin{minipage}[t]{\textwidth}
        \begin{minipage}[b]{0.8\textwidth}
          \textbf{Release()}:
          \begin{algorithmic}[1] \small
            \State glb $:=^{\sf R}$ r + 2
          \end{algorithmic}
        \end{minipage}
      \end{minipage}
      \caption{Sequence lock in RC11-RAR}
      \label{seqlock_imp}
      \vspace{-1em}
    \end{minipage}\hfill
        \begin{minipage}[b]{0.48\textwidth}
      \textbf{Init:} \ \ nt = 0, \ \ sn = 0\\[6pt]
      \begin{minipage}[t]{\columnwidth}
 \begin{minipage}[b]{0.80\textwidth}
   \textbf{Acquire()}:
        \begin{algorithmic}[1] \small
          \State m\_t $\leftarrow$ $\kwfai^{\sf RA}$(nt)
          \State  \textbf{do}
             s\_n $\leftarrow^{\sf A}$ sn
             \textbf{until} m\_t = s\_n
        \end{algorithmic}
      \end{minipage}
      \end{minipage}
      \\[8pt]     
      \begin{minipage}[t]{\columnwidth}
        \begin{minipage}[b]{0.80\textwidth}
          \textbf{Release()}: 
        \begin{algorithmic}[1] \small
          \State sn $:=^{\sf R}$ s\_n + 1
        \end{algorithmic}
        \end{minipage}
      \end{minipage}
      \caption{Ticket lock in RC11-RAR}
      \label{ticketlock_imp}
          \vspace{-1em}
    \end{minipage}
  \end{figure}

\begin{proposition}
  For synchronisation-free clients, there exists a forward simulation
  between the abstract lock object and the sequence
  lock. 
\end{proposition}
\subsection{Ticket Lock}
\label{sec:ticket-lock}
Our second refinement example (\reffig{ticketlock_imp}) is the ticket
lock.  The ticket lock has two shared variables $nt$ (next ticket) and
$sn$ (serving now). Invocation of {\bf Acquire} loads the next
available ticket into a local register ($m\_t$) and increases the
value of $nt$ by one using a fetch-and-increment ($\kwfai$)
operation. It then enters a busy loop and reads $sn$ until it sees its
own ticket value in $sn$ before it can enter its critical section.

If the read on line 2 of the {\bf Acquire} operation reads from a
write whose value is equal to the value of $m\_t$, then the
lock is acquired. Therefore we will need to show that if this
situation arises, the abstract lock object can also take a step and
successfully acquire the lock. We consider the $\kwfai$ operation on
line 1 and the read on line 2 if it reads a value that is not equal to
$m\_t$ to be a stuttering step. We prove that each of the
stuttering and non-stuttering steps preserves the simulation
relation. Similar to the previous example, the {\bf Release} operation
consists of only one releasing write to variable $sn$ and it is
straightforward to show that this operation refines the abstract
release operation. This proof has been mechanised in Isabelle/HOL. 

\begin{proposition}
  For synchronisation-free clients, there exists a forward simulation
  between the abstract lock object and the ticket
  lock. 
\end{proposition}



\newcommand{\supp}{{\it supp}}
\newcommand{\nVal}{{\it nVal}}
\newcommand{\nNxt}{{\it nNxt}}
\newcommand{\nTop}{{\it nTop}}
\newcommand{\lastVal}{{\it lastVal}}
\newcommand{\lastPush}{{\it lastPush}}
\newcommand{\usedAddr}{{\it usedAddr}}
\newcommand{\pushedAddr}{{\it pushedAddr}}
\newcommand{\unmatchedPush}{{\it unmatchedPush}}
\newcommand{\injective}{{\it injective}}

\newcommand{\nxtRel}{{\it nxtRel}}

\section{Concurrent Data Structures}
\label{sec:view-based-relax}

The modelling approach in \refsec{sec:abstr-object-semant} provides a
flexible framework for specifying abstract objects and enables
specification of synchronised and unsynchronised method calls (as
detailed in \reffig{fig:publication}). We now show that the framework
also provides a mechanism for specifying concurrent objects whose
linearisation order is different from real-time order (cf.
\cite{DBLP:journals/pacmpl/RaadDRLV19,DongolJRA18,DBLP:conf/esop/KrishnaEEJ20,DBLP:journals/pacmpl/EmmiE19}).
Such specifications are a natural fit with the weak memory operational
semantics where new writes may be introduced in the ``middle'' of
modification
order~\cite{DBLP:conf/ppopp/DohertyDWD19,Dolan:2018:LDRF,DBLP:conf/popl/KangHLVD17}. 
 


\subsection{Specifying C11-Style Concurrent Data Structures}
\label{sec:specifying-c11-style}





As with ordinary read, write and RMW operations, there
are two separate aspects to consider.
\begin{enumerate}
\item The introduction of operations into modification order. As with
  the operational C11
  semantics~\cite{ECOOP20,DBLP:conf/ppopp/DohertyDWD19}, we allow
  operations to be introduced into the middle of modification order
  (under some restrictions on thread views).
\item The release-acquire synchronisation guarantees induced by the
  operations. Concurrent data structures are to be implemented as
  concurrent libraries, and hence induce inter-thread synchronisation
  across both the library and the client.
\end{enumerate}

As with writes, timestamps may be used to define a linear order of the
method calls of the object in question. However, \emph{different}
threads may have \emph{different} views of the object being
implemented, and this phenomenon must be supported by the
object specification. 
To exemplify our approach,  a queue and a stack specification are.
 given below



\begin{example}[Queue specification]
  \label{ex:queue}
  Consider a timeline corresponding to the state of a shared queue
  as depicted below:
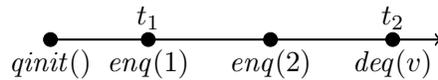
\begin{center}
  \begin{tikzpicture}[scale=0.65]
    \draw[thick,->] (-1,0) -- (7,0);
        
    \coordinate (init) at (-1,0);
    \coordinate (enq1) at (1,0);
    \coordinate (enq2) at (3.5,0);
    \coordinate (deq2) at (6,0);
    
    \draw (init) [fill] circle (4pt)
    node[below]  {$\initq()$};

    \draw (enq1) [fill] circle (4pt)
    node[below]  {$\enq(1)$}
    node[above]  {$t_1$};
    
    \draw [black] (enq2) [fill] circle (4pt)
    node[below]  {$\enq(2)$};
    
    \draw [black] (deq2) [fill] circle (4pt)
    node[below]  {$\deq(v)$} 
    node[above]  {$t_2$};

    
  \end{tikzpicture}
\end{center}
where the viewfront of threads $t_1$ and $t_2$
are depicted above the timeline and the abstract operations executed
below the timeline. The state can be generated by executing
the operations in the following temporal order:
\begin{enumerate}
\item Thread $t_2$ executes $\enq(2)$, and hence $t_2$'s view of the
  queue is that of $\enq(2)$.
\item Thread $t_1$ executes $\enq(1)$, but since it has not yet
  ``seen'' $\enq(2)$, it may choose a timestamp that is smaller than
  the timestamp of $\enq(2)$ for $enq(1)$.
\item Thread $t_2$ executes a dequeue, $\deq(v)$, for some yet to be
  determined value $v$. Since $t_2$ also executed $\enq(2)$, the
  timestamp of the $\deq(v)$ must be larger than that of
  $\enq(2)$. 
\end{enumerate}

One option for the execution to continue in a sensible manner is to
ensure that the effect of the dequeue is communicated to all other
threads so that their views of the object are updated. Thus, if $t_1$
were to perform a new method call, this method call would appear after
$\deq(v)$ (with $v = 1$). However, imposing such a communication
requirement in the abstract specification induces additional (and
potentially unnecessary) synchronisation in an implementation. For
instance, a sensible continuation of the timeline in above would be
for thread $t_1$ to introduce additional enqueue operations between
$\enq(1)$ and $\deq(v)$ (with $v = 1$). In fact, the only
continuations that are disallowed is the introduction of new dequeue
operations between $\enq(1)$ and $\deq(v)$ (with $v = 1$) and new enqueue operations
before $\enq(1)$.

Our formal specification builds on this idea. We assume a predicate,
$\matchedTS$, that stores pairs of timestamps corresponding to
enqueues and dequeues. For any $(ts, ts') \in \matchedTS$, the
timestamp $ts$ corresponds to an enqueue that is removed from the
queue by the dequeue with timestamp $ts'$. To ensure
first-in-first-out behaviours, for any pair
$(ts, ts') \in \matchedTS$, it is illegal for a dequeue to be
introduced between $ts$ and $ts'$. Moreover, when a dequeue matches an
operation, it must match with the first unmatched enqueue that has a
smaller timestamp than the dequeue (or empty if no such enqueue
exists).  Below, we present a specification of enqueue and dequeue
operations. 
We let
$\supp(R) = \dom(R) \cup \ran(R)$ for a relation $R$.\medskip

\noindent $\small \inference[{\sc Enq-Base}] {
    \tst(\alpha.\tview_t(q)) < \ts' \qquad     \writes' = \alpha.\writes \cup \{(a, \ts')\} \\
    (\forall (\mathit{ww}, \mathit{tt}) \in \alpha.\writes.\
    \ww \in Enq \wedge \tts > \ts'  \imp \mathit{tt} \notin \dom(\alpha.\matchedTS))
    \\
    (\forall (\ww, \tts) \in \alpha.\writes.\ \tts > \ts' \imp ww \noteq q.deq_{\it empty}) \qquad 
    \tview' = \alpha.\tview_t[q := (a, \ts')] 
  }
  {\alpha\ \Strans{a, \ts'}_{t}\  \alpha[\writes \asgn \writes', \tview_t \asgn \tview'] }
$ \medskip

\noindent $\small \inference[{\sc Enq-Rlx}] {
    a = q.enq_u 
    \qquad 
    \alpha\ \Strans{a, \ts'}_{t}\  \alpha' 
  }
  {\alpha, \gamma\  \strans{a}_{t}\  \alpha', \gamma }
\qquad  \inference[{\sc Enq-Sync}] {
    a =q.enq_u^{\sf R} \qquad
\alpha\  \Strans{a, \ts'}_{t}\  \alpha' \\ 
    \mview' = \alpha'.\tview_t \cup \gamma.\tview_t 
  }
  {\alpha, \gamma\  \strans{a}_{t}\  \alpha'[\mview_{(a,\ts')} \asgn \mview'], \gamma }
  $ \medskip
  
\noindent $\small  \inference[{\sc Deq-Base}] {
    \max(\tst(\alpha.\tview_t(q)), \ts) < \ts' \\ \writes' = \alpha.\writes \cup \{(a, \ts')\}\qquad 
    (w, \ts) \in \alpha.\writes \qquad \ts \notin \dom(\alpha.\matchedTS)  
    \\ 
    (\forall (\ww, \tts) \in \alpha.\writes.\ ww \in Enq \wedge \tts < \ts \imp \tts \in \dom(\alpha.\matchedTS)) \\
    (\forall \tts' \in \ran(\alpha.\matchedTS).\ ts' > \tts')
   \qquad
    \tview' = \alpha.\tview_t[q := (a, \ts')]
  }
  {\alpha\  \Strans{a, \ts', w, ts}_{t}\
   \alpha [\writes \asgn \writes', \tview_t \asgn \tview', \matchedTS := \matchedTS \cup \{(\ts, \ts')\}]}\\
$\medskip

\noindent 
$ \small \inference[{\sc Deq-NE-Rlx}] {
    a = q.deq^{\sf [A]}_u \quad u \neq {\it empty} \qquad 
    \alpha\  \Strans{a, \ts', w, ts}_{t}\
    \alpha' \qquad w = q.\enq^{\sf [R]}_u \quad (w, a) \notin {\sf R} \times {\sf A} 
  }
  {\alpha, \gamma\  \strans{a}_{t}\  \alpha', \gamma}
$\medskip

\noindent$ \small \inference[{\sc Deq-NE-Sync}] {
    a = q.deq^{\sf A}_u \qquad u \neq {\it empty} \qquad \alpha\  \Strans{a, \ts', w, ts}_{t}\
   \alpha' \qquad w = q.enq^{\sf R}_u  \\
    \tview' = \alpha'.\tview_t  \otimes \alpha.\mview_{(w,\ts)}
    \qquad
    \ctview' = \gamma.\tview_t  \otimes \alpha.\mview_{(w,\ts)}
  }
  {\alpha, \gamma\  \strans{a}_{t}\  \alpha'[\tview_t \asgn \tview'
], \gamma[\tview_t := \ctview']}\\
$\medskip

\noindent
$ \small \inference[{\sc Deq-Emp}] {
    a = q.deq_{\it empty} \qquad \tst(\alpha.\tview_t(q)) < \ts' \qquad \writes' = \alpha.\writes \cup \{(a, \ts')\}\\
    (\forall (\ww, \tts) \in \alpha.\writes.\ \tts < \ts'
      \imp  \tts \in \supp(\sigma.\matchedTS) \lor \ww = q.deq_{\it empty})
\\
    \tview' = \alpha.\tview_t[q := (a, \ts')] 
  }
  {\alpha, \gamma\  \strans{a}_{t}\  \alpha[\writes \asgn \writes',
    \tview_t \asgn \tview'], \gamma }
$ \bigskip

Rules {\sc Enq-Base} and {\sc Deq-Base} define the basic rules for
enqueues and dequeues, factoring out the common aspects of relaxed and
synchronising operations. {\sc Enq-Base} helps determine an
appropriate timestamp for a new (enqueue) operation and advances the
thread view of the executing thread to this timestamp. For a relaxed enqueue,
{\sc Enq-Rlx}, no other state changes are necessary. However, for an acquiring enqueue, the modification view of the new operation must be updated as defined by {\sc Enq-Sync}. 

{\sc Deq-Base} is similar to {\sc Enq-Base}, but it also makes the
enqueue that the dequeue matches with explicit. This is used in rule
{\sc Deq-NE-Sync} to update thread views of both the library and the
client to be consistent with the corresponding (releasing) enqueue operation.

\medskip

\end{example}

\begin{example}[Stack specification]
  \label{ex:stack}
  The specification of a stack is designed
  to ensure last-in-first-out order. We motivate the main ideas of the
  specification below. Consider the following timeline corresponding to the state
  of a shared stack: 
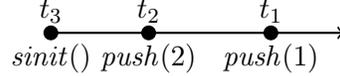
\begin{center}
  \begin{tikzpicture}[scale=0.65]
    \draw[thick,->] (-1,0) -- (5,0);
        
    \coordinate (init) at (-1,0);
    \coordinate (enq1) at (1,0);
    \coordinate (enq2) at (3.5,0);
    \coordinate (deq2) at (6,0);

    \draw (init) [fill] circle (4pt)
    node[below]  {$\inits()$}
    node[above]  {$t_3$};

    \draw (enq1) [fill] circle (4pt)
    node[below]  {$\push(2)$}
    node[above]  {$t_2$};
    
    \draw [black] (enq2) [fill] circle (4pt)
    node[above] {$t_1$}
    node[below]  {$\push(1)$};
    
  \end{tikzpicture}
\end{center}
In this state, 
thread $t_1$'s viewfront is at $\push(1)$, thread $t_2$'s viewfront is
at $\push(2)$ and thread $t_3$'s viewfront is at $\inits()$. Consider
a $\pop$ operation performed by each of the threads --- these
operations must be consistent with their viewfronts. For the state
above, a pop executed by $t_3$ may return either empty, $1$ or $2$, a
pop executed by thread $t_2$ may return either $2$ or $1$, and a pop
executed by thread $t_1$ must return $1$.  

Suppose $t_3$ performs a $pop$, and that the pop returns the value $2$. In this case, the state would be updated as follows:
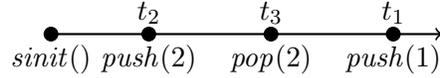
\begin{center}
  \begin{tikzpicture}[scale=0.65]
    \draw[thick,->] (-1,0) -- (7,0);
        
    \coordinate (init) at (-1,0);
    \coordinate (enq1) at (1,0);
    \coordinate (deq2) at (3.5,0);
    \coordinate (enq2) at (6,0);
    
    
    
    \draw (init) [fill] circle (4pt)
    node[below]  {$\inits()$};


    \draw (enq1) [fill] circle (4pt)
    node[below]  {$\push(2)$}
    node[above]  {$t_2$};
    
    \draw [black] (enq2) [fill] circle (4pt)
    node[above] {$t_1$}
    node[below]  {$\push(1)$};
    
    \draw [black] (deq2) [fill] circle (4pt)
    node[below]  {$\pop(2)$} 
    node[above]  {$t_3$};

    
  \end{tikzpicture}
\end{center}
Note that $t_3$'s viewfront must include $push(2)$, so $pop(2)$ must
occur after $push(2)$. However, $pop(2)$ must not occur after
$push(1)$; otherwise the popped value would be inconsistent with the
history of a sequential stack since the history
$\push(2)\, \push(1)\, \pop(2)$ is invalid.

Now consider an alternative execution starting from the first state
shown above, where thread $t_1$ performs a pop. The post state following this execution is given below: 
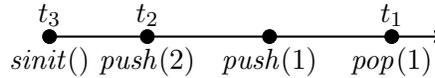
\begin{center}
  \begin{tikzpicture}[scale=0.65]
    \draw[thick,->] (-1,0) -- (7,0);
        
    \coordinate (init) at (-1,0);
    \coordinate (enq1) at (1,0);
    \coordinate (deq2) at (3.5,0);
    \coordinate (enq2) at (6,0);

    \draw (init) [fill] circle (4pt)
    node[below]  {$\inits()$}
    node[above]  {$t_3$};

    \draw (enq1) [fill] circle (4pt)
    node[below]  {$\push(2)$}
    node[above]  {$t_2$};    

    \draw [black] (deq2) [fill] circle (4pt)
    node[below]  {$\push(1)$};

    \draw [black] (enq2) [fill] circle (4pt)
    node[below]  {$\pop(1)$} 
    node[above]  {$t_1$};
  \end{tikzpicture}
\end{center}
Now suppose $t_2$ performs a $pop$. This pop must return the value $2$
since $\push(1)$ has already been popped and $t_2$'s viewfront
includes $\push(2)$. Moreover, since $\push(1)$ and $\pop(1)$ are
already matched, it would be illegal for the pop by $t_2$ to be
introduced between $\push(1)$ and $\pop(1)$. However, it is acceptable
to introduce the pop before $\push(1)$ or after $\pop(1)$ since this
histories $\push(2)\, \pop(2)\, \push(1)\, \pop(1)$ and
$\push(2)\, \push(1)\, \pop(1)\, \pop(2)$ are both valid.

These ideas are captured by the semantics below. Rules {\sc Push-Base} and {\sc Pop-Base} capture the basic ordering
requirements and thread view updates. Rules {\sc Push-Rlx} and {\sc
  Push-Sync} define relaxed and (releasing) pushes, respectively,
while rules {\sc Pop-Rlx} and {\sc Pop-Sync} define unsynchronised and
synchronised pops, respectively.  

\medskip

\noindent
$\small \inference[{\sc Push-Base}] {
    \tst(\alpha.\tview_t(q)) < \ts' \qquad     \writes' = \alpha.\writes \cup \{(a, \ts')\} \\
    (\forall 
    (tt, tt') \in \alpha.\matchedTS.\ \ts' < tt \lor ts' > tt')
    \qquad 
    \tview' = \alpha.\tview_t[q := (a, \ts')]
  }
  {\alpha\  \Strans{a, \ts'}_{t}\  \alpha[\writes \asgn \writes', \tview_t \asgn \tview']}
  $ \smallskip
  
\noindent
$\small \inference[{\sc Push-Rlx}] {
    a =  q.push(u) \qquad \alpha\  \Strans{a, \ts'}_{t}\  \alpha' 
  }
  {\alpha, \gamma\  \strans{a}_{t}\  \alpha', \gamma }
\quad \inference[{\sc Push-Sync}] {
    a =  q.push(u)^{\sf R} \qquad \alpha\  \Strans{a, \ts'}_{t}\  \alpha' \\ 
    \mview' = \tview' \cup \gamma.\tview_t 
  }
  {\alpha, \gamma\  \strans{a}_{t}\  \alpha'[\mview_{(a,\ts')} \asgn \mview'], \gamma }
  $\smallskip
  
\noindent
$\small \inference[{\sc Pop-Base}] {
\alpha\  \Strans{a, \ts'}_{t}\  \alpha' \qquad
    (w, \ts) \in \alpha.\writes \qquad \ts \notin \dom(\alpha.\matchedTS)
    \qquad \ts < \ts' \\
    (\forall (ww, \ts'') \in \alpha.\writes.\ ww \in Push \wedge \ts < \ts'' < \ts'
    \imp \ts'' \in \dom(\alpha.\matchedTS))
  }
  {\alpha\  \Strans{a, \ts', w, \ts}_{t}\  \alpha'[\matchedTS := \matchedTS \cup \{(\ts, \ts')\}]}
  $
  \smallskip
  
\noindent
  $\small  \inference[{\sc Pop-NE-Rlx}] {
    a = q.pop^{\sf [R]}_u \quad u \neq {\it empty} \quad \alpha\  \Strans{a, \ts', w, \ts}_{t}\  \alpha' \quad w = q.push^{\sf [A]}_u \quad (w, a) \notin {\sf R} \times {\sf A} \\
  }
  {\alpha, \gamma\  \strans{a}_{t}\  \alpha', \gamma}
  $\smallskip

  \noindent
$\small  \inference[{\sc Pop-NE-Sync}] {
    a = q.pop(u)^{\sf A} \qquad u \neq {\it empty} \qquad \alpha\  \Strans{a, \ts', w, \ts}_{t}\  \alpha' \qquad w = q.push(u)^{\sf R}
    \\
    \tview' = \alpha'.\tview_t  \otimes \alpha.\mview_{(w,\ts)}
    \qquad 
    \ctview' = \gamma.\tview_t  \otimes \alpha.\mview_{(w,\ts)}
  }
  {\alpha, \gamma\  \strans{a}_{t}\  \alpha'\left[
    \begin{array}[c]{@{}l@{}}
      \tview_t \asgn \tview'
    \end{array}
\right], \gamma[\tview_t := \ctview']}
$ \smallskip

\noindent
$\small \inference[{\sc Pop-Emp}] {
    a = q.pop^{[{\sf A}]}_{\it empty} \qquad \tst(\alpha.\tview_t(q)) < \ts' \qquad \writes' = \alpha.\writes \cup \{(a, \ts')\}\\
    (\forall (\ww, \tts) \in \alpha.\writes.\ \tts < \ts'
      \imp  \tts \in \sup(\sigma.\matchedTS) \lor \ww = q.pop^{[{\sf A}]}_{\it empty})
\\
    \tview' = \alpha.\tview_t[q := (a, \ts')] 
  }
  {\alpha, \gamma\  \strans{a}_{t}\  \alpha[\writes \asgn \writes',
    \tview_t \asgn \tview'], \gamma }
$  
\end{example}
We contrast this specification with the prior declarative semantics of
Raad et al.~\cite{DBLP:journals/pacmpl/RaadDRLV19}, which defines
semantics for complete executions.  Note that Raad et
al.~\cite{DBLP:journals/pacmpl/RaadDRLV19} also describe weak stack
specifications; but their notion of weakness relaxes the total order
of operations into a partial order. In our model, the operations are
still totally ordered, but this total order may be different from the
real-time order of operations, thus represents a relaxation of
abstract specifications as used in proofs of
linearizability~\cite{HeWi90,DongolD15}.

With this specification, it becomes possible to make the proof outline
described in \reffig{fig:mp-proof} more precise. In particular, we can
prove the following property, which we have verified using our
Isabelle/HOL mechanisation. 

\begin{proposition}
  \label{prop:stack-mp}
  The proof outline in \reffig{fig:mp-proof} is
  valid.
\end{proposition}

Note that the variable $d$ in \reffig{fig:mp-proof} is a client
variable, whereas the stack operations are part of an abstract
library. The synchronisation guarantees provided by the stack library
are sufficient to guarantee message passing in the client. Next, in
\refsec{sec:treiber-stack}, we show that such a specification can
indeed be implemented. In particular, we show that the Treiber stack
adapted for C11, as described by Doherty et al \cite{ifm18} satisfies
the stack specification from \refex{ex:stack}, i.e., the
implementations of the push and pop operations in \refalg{alg:treiber}
provide sufficient synchronisation to provide client-side message
passing.

\subsection{C11 Treiber Stack}
\label{sec:treiber-stack}

 \setlength{\textfloatsep}{6pt}

\begin{algorithm}[!t] 
  \caption{Treiber stack with relaxed and release-acquire memory accesses}
  \label{alg:treiber}
  \begin{varwidth}[t]{0.45\columnwidth}
    \begin{algorithmic}[1]
      \Procedure{\textsc{Init}}{}
      \State Top := null; 
      \EndProcedure
      \algstore{tml-ctr}
    \end{algorithmic}
    
    \begin{algorithmic}[1]
      \algrestore{tml-ctr} \small
      \Procedure{\textsc{Push}$^{\sf R}$}{v}
      \State n := \kwnew(node); 
      \State n.val := v ; \label{write-val}
      \Repeat 
      \State top $\gets^{\sf A}$ Top ;  \label{push-acquire}
      \State n.nxt := top  ; \label{write-nxt}
      \Until {$\kwcas^{\sf R}$(\& Top, top, n) } \label{push-release}
      \EndProcedure
      \algstore{tml-ctr}
    \end{algorithmic}

  \end{varwidth}
  \hfill
  \begin{varwidth}[t]{0.5\columnwidth}
    \begin{algorithmic}[1]
      \algrestore{tml-ctr}
      \Function{\textsc{Pop}$^{\sf A}$}{} 
      \Repeat 
      \Repeat \label{spin1}
        \State top $\gets ^{\sf A}$ Top ; \label{pop-acquire} \label{spin2}
      \Until {top $\neq$ null} ; \label{spin3}
      \State nTop $\gets$ top.nxt ;
      \Until {$\kwcas^{\sf R}$(\& Top, top, nTop)} \label{pop-release}
      \State \Return top.val ;
      \EndFunction 
    \end{algorithmic}
  \end{varwidth}

\end{algorithm}

The example we consider (see \refalg{alg:treiber}) is the Treiber
Stack \cite{Tre86} (well-studied in a SC setting, but not in a weak
memory one), executing in the RC11~\cite{DBLP:conf/pldi/LahavVKHD17} memory model.  
Due to weak memory effects, the events under consideration, including
method invocation and response events are partially ordered
\cite{AlglaveMT14,DBLP:conf/popl/LahavGV16,DBLP:conf/vmcai/DokoV16,DBLP:conf/pldi/LahavVKHD17,DBLP:conf/popl/BattyDG13}. 
In \refalg{alg:treiber}, all accesses to 
Top occur in weak memory. In particular, any read of Top (lines
\ref{push-acquire}, \ref{pop-acquire}) reading from a write to Top
(lines~\mbox{\ref{push-release}, \ref{pop-release}}) induces
\emph{happens-before order} from the write to the read. Any read of
Top via a failed CAS is relaxed (see \reffig{fig:comm-sem}).

For simplicity, like Doherty et al~\cite{ifm18} we assume that the
\textsc{Pop} operation blocks by spinning instead of returning
empty. Doherty et al show that the Treiber Stack in
\refalg{alg:treiber} satisfies a property known as {\em causal
  linearizability}, which in turn guarantees {\em compositionality}:
any history containing executions of two causally linearizable objects
will itself be causally linearizable. However, nothing is known about
the client-side guarantees of the algorithm (namely, whether what
client-side synchronisation guarantees are available).

In this section, we address this shortcoming by proving refinement
between the implementation (\refalg{alg:treiber}) and the stack
specification described in \refex{ex:stack}. In particular, we show
that the {\sc Push$^{\sf R}$}(v) and {\sc Pop$^{\sf A}$} operations
respectively implement the synchronising push and pop operations in
\refex{ex:stack}.

\begin{algorithm}[t]
  \caption{Translated model and proof outline}
  \label{alg:treiber-PO}
  \begin{varwidth}[t]{0.5\columnwidth}
    
    \begin{algorithmic}[1]
      \setcounter{ALG@line}{2} \small
      \Procedure{\textsc{Push}$^{\sf R}$}{v}
      \Statex \quad
      $\color{green!40!black}
          \{I \wedge \forall a \notin \usedAddr.\ [a = 0]_t^L\}$
          \State ${\bf choose}\  a\ {\bf where}\ a, a+1 \notin \usedAddr:$
          \Statex \qquad  $\nVal_t, \nNxt_t, \usedAddr :=$
          \Statex \qquad\qquad  $a, a + 1, \usedAddr \cup \{a, a+1\}$; 
      \Statex \quad
      $\color{green!40!black}
          \{I \wedge [\nVal_t = 0]_t^L \wedge [\nNxt_t = 0]_t^L\}$
      \State $\nVal_t := v$ ; \label{write-val}
      \Statex \quad
      $\color{green!40!black}
          \{I \wedge [\nVal_t = v]_t^L \wedge [\nNxt_t = 0]_t^L\}$
      \Repeat 
      \Statex \quad\quad\ \ 
      $\color{green!40!black}
          \{I \wedge [\nVal_t = v]_t^L \wedge \exists v. [\nNxt_t = v]_t^L\}$
      \State $top_t \gets^{\sf A} \Top$ ;  \label{push-acquire}
      \Statex \quad\quad
      $\color{green!40!black}
          \{I \wedge [\nVal_t = v]_t^L \wedge \exists v. [\nNxt_t = v]_t^L\}$
      \State $\nNxt_t := top_t$  ; \label{write-nxt}
      \Statex \quad\quad
      $\color{green!40!black}
          \{I \wedge [\nVal_t = v]_t^L \wedge [\nNxt_t = top_t]_t^L\}$
          \Until {$\kwcas^{\sf R}(\Top, top_t, \nVal_t)$ }  \label{push-release}
      \EndProcedure
      \algstore{tml-ctr}
    \end{algorithmic}

  \end{varwidth}
  \hfill
  \begin{varwidth}[t]{0.5\columnwidth} \small
    \begin{algorithmic}[1]
      \algrestore{tml-ctr}
      \Function{\textsc{Pop}$^{\sf A}$}{} 
      \Repeat 
      \Repeat \label{spin1}
      \Statex \qquad\qquad  $\color{green!40!black}
      \{I \wedge CO_{pop}\}$
      \State $top_t \gets ^{\sf A} Top$ ; \label{pop-acquire} \label{spin2}
      \Statex \qquad\qquad $\color{green!40!black}
      \{I \wedge CO_{pop} \wedge DO_{pop}\}$
      \Until {$top_t \neq null$} ; \label{spin3}
      \Statex \qquad\ \ $\color{green!40!black}
      \{I \wedge CO_{pop} \wedge DO_{pop}\}$
      \State $\nTop_t \gets (top_t + 1)$ ;
      \Statex \qquad\ \  $\color{green!40!black}
      \{I \wedge CO_{pop} \wedge DO_{pop}\}$
      \Until {$\kwcas^{\sf R}(\Top, top_t, \nTop_t)$} \label{pop-release}
      \Statex \quad\ \ $\color{green!40!black}
      \{I \wedge \exists v . [top_t = v]_t^L\}$
      \State $rval_t \gets top_t$ ;
      \EndFunction 
    \end{algorithmic}
  \end{varwidth}

  \smallskip \smallskip
  \small
  
  where ${\it TopVals}$ is the set of all values for all writes to
  $\Top$ in the given weak memory state, and
  \begin{small}
  \begin{align*} \small
    I & \small= \refeq{inv-1} \wedge \refeq{inv-2} \wedge \refeq{inv-3} \wedge \refeq{inv-4}
        \wedge \refeq{inv-5} \wedge \refeq{inv-6} 
        \\\small
    CO_{pop} & =
               \begin{array}[t]{@{}l@{}}
                 (\exists v. [Top = v]^L_t) \\
                 \wedge (\forall a \in {\it TopVals}.\  (\exists v . \langle \Top = a\rangle^L[a = v]_t^L) \wedge (\exists v . \langle \Top = a\rangle^L[(a + 1) = v]_t^L))
               \end{array}
    \\
    \small DO_{pop} & = (\exists v .\ [top_t = v]_t^L) \wedge (\exists v .\ [(top_t + 1) = v]_t^L)
  \end{align*}
\end{small}
\end{algorithm}

\smallskip \noindent {\bf Modelling the Treiber Stack.} We track a set
of used addresses, $\usedAddr$. Specifically, each address in $\usedAddr$ is an
address containing the value ($n.\fval$) and next field ($n.\fnxt$) of
the node that a thread is trying to push, or has already been pushed
onto the stack. The variable $\Top$ is a (relaxed) pointer --- the
value of each write to $\Top$ is an address from $\usedAddr$ or null. The $\fval$
and $\fnxt$ fields of each node are modelled by (relaxed) writes to
addresses in $\usedAddr$. 



For the push, at line 4, two previously unused addresses are selected
and inserted into $\usedAddr$; these addresses is recorded in local
variable $\nVal_t$ and $\nNxt_t$ for thread $t$. We assume that the
new node constructor allocates two consecutive addresses, and hence
assume $\nNxt_t = \nVal_t + 1$. We also assume that both variables are
initialised with value $0$. In our development, $\Top$ will point to
an address holding the value, say $\nVal$, and the next node is
defined as the value of the last write to address $\nVal + 1$.  For
line 5, we introduce a relaxed write to the address corresponding to
$\nVal_t$ with value $v$. Line~7 corresponds to an acquiring read of
$\Top$, line 8 updates the $\fnxt$ field, and line 10 executes a
releasing $\kwcas$ on $\Top$. Similarly, for the pop, at line 13, we
introduce an acquiring read of $\Top$ into local variable $top_t$. At
line 15, we generate a read of the address $top_t + 1$ (which, as
describe above, dereferences the next field), and store the value of
this address in local variable $\nTop_t$. At line 16, we perform a
releasing $\kwcas$ on $\Top$.


\smallskip
\noindent {\bf Treiber Stack invariants.} We use an
auxiliary variable $\pushedAddr$ to keep track of the addresses that
have been pushed on to the stack. We set $\pushedAddr$ to
$\pushedAddr \cup \{\nVal_t\}$ at line 9 and to
$\pushedAddr \setminus \{top_t\}$ at line 16 whenever the
corresponding $\kwcas$ operations succeed.  Given a library state
$\beta$, the addresses that have been pushed onto the stack (in order)
are thus given by $\nxtRel_\beta$ (defined below), recalling that we
use address $a+1$ to store writes to the next field. Thus:
\begin{align*}
  \lastVal_\beta(x) & = \wrval(\last(\beta.\writes, x))
  \\
  \nxtRel_{\mem, \beta} & = \{(a,b) \mid  a \in \mem.\pushedAddr \wedge
                  b = \lastVal_\beta(a + 1) \}^+
\end{align*}
Thus, $\nxtRel_{\mem, \beta}$ is the transitive closure of the set
that maps each pushed address (as recorded in $\mem$) to the
address of the next node. The first component of our global invariant
ensures that $nxtRel_\beta$ is a total order with $\Top$ as the
minimal element ($\inf$) and $null$ as the maximal element ($\sup$).
\begin{align}
    \label{inv-1}
    & totalOrder(nxtRel_\beta) \wedge \inf(nxtRel_\beta) = {\it lastVal}_\beta(Top) \wedge \sup(nxtRel_\beta) = null
\end{align}



\noindent
The next two components of the invariant state properties about
$\Top$, and links these values to the auxiliary state components
$\usedAddr$ and $pushedAddr$.
\begin{align}
  \label{inv-2}
    & \forall  w \in \beta.\writes_{| \Top}.\ w \in \W_{\sf R} \wedge (\wrval(w)  \neq  null  \imp   \wrval(w)  \in  \mem.\usedAddr)
  \\
  \label{inv-3}
  & {\it lastVal}_\beta(Top)  \neq  null  \imp   {\it lastVal}_\beta(Top)  \in  \mem.\pushedAddr
\end{align}
Invariant \refeq{inv-2} states writes to $\Top$ are releasing and
non-null writes are recorded in $\mem.\usedAddr$, while \refeq{inv-3}
states that whenever the last value of $\Top$ is non-null, this value
is a currently pushed address.\footnote{Invariant \refeq{inv-3} could
  be strengthened so that we have
  ${\it lastVal}_\beta(a) \in \mem.\pushedAddr$ for all
  $a \in \dom(\nxtRel_{\mem, \beta})$, but is not needed for our
  verification. }
Finally, we have a set of simple invariants over $\usedAddr$ and
$\pushedAddr$.
\begin{align}
  \label{inv-4}
    & \mem.\pushedAddr  \subseteq \mem.\usedAddr
  \\
  \label{inv-5}
 & \forall a_1, a_2 \in \mem.\pushedAddr.\  a_1 \neq a_2  \imp  \lastVal_\beta(a_1 + 1)  \neq  \lastVal_\beta(a_2 + 1)
  \\
  \label{inv-6}
    & \forall  a \in \mem.\pushedAddr.\  \lastVal_\beta(a + 1)  \neq  null  \imp  \lastVal_\beta(a + 1) \in \mem.\pushedAddr
\end{align}
Invariant \refeq{inv-4} states that all pushed addresses are used
addresses, \refeq{inv-5} states that the next pointers of different
pushed addresses are different and \refeq{inv-6} states that if the
value of the next pointer of a pushed address is non-null, then the
node corresponding to the next pointer is a pushed address.

\smallskip \noindent {\bf Simulation relation.} We now describe the
simulation relation used to prove refinement between the Treiber Stack
and the stack specification in \refsec{ex:stack}.  A key component of
the relation is an injective function, $f$, mapping nodes
corresponding to pushed addresses to unmatched push operations of the
abstract specification. Recall from \refex{ex:stack} that pushes are
matched if they have been subsequently popped. The refinement relation
$R_{\it TS}((als, \gamma_A, \alpha), (ls, \gamma_C, \beta)) = \exists f.\
\refeq{ref-2} \wedge \refeq{ref-3a} \wedge \refeq{ref-3} \wedge
\refeq{ref-3b} \wedge \refeq{ref-5} \wedge \refeq{ref-6} \wedge
\refeq{ref-7}$.

The first component ensures injectivity of $f$ 
and that 
the corresponding operation is a pop that
returns empty.
\begin{align}
  \label{ref-2}
  & \injective(f) 
    \wedge \exists ts.\ f(null) = (pop(empty), ts) 
\end{align}
We assume $\unmatchedPush_\alpha$ denotes the set of unmatched push
operations in the abstract state $\alpha$. The next two conditions
describe the relationship between pushed addresses in the concrete
state to the unmatched pushes:
\begin{align}
    \label{ref-3a}
  &       
    \begin{array}[c]{@{}l@{}l@{}}
      \forall  a \in \mem.\pushedAddr .\ \ & f(a) \in  \unmatchedPush_\alpha   \wedge val_\alpha (f(a)) = {\it lastVal}_\beta(a)  \\
      & {} \wedge \tst(f(a)) > \tst(f (\lastVal_\beta(a+1))) 
    \end{array}
  \\
    \label{ref-3}
  & \begin{array}[c]{@{}l@{}}
      \forall  a \in \mem.\pushedAddr,  op \in \unmatchedPush_\alpha.\ \\
      \qquad op \neq f(a) \wedge op \neq f (\lastVal_\beta(a+1)) \imp   \\
      \qquad  \tst(op) < \tst(f (\lastVal_\beta(a+1)) \lor \tst(op) > \tst(f(a))
    \end{array}
\end{align}
Condition \refeq{ref-3a} states that for each pushed address, $f(a)$
is indeed an unmatched push with the same values. Moreover, the
timestamp of the push corresponding $f(a)$ is larger than the
timestamp of the push corresponding to the next node of $a$. Condition
\refeq{ref-3} 
ensures that the timestamp of $op$ is not in between that of $f(a)$
and the immediately next node of $f(a)$.

Suppose $\lastPush_\alpha$ returns the push in $\alpha$ with the
largest timestamp. The next condition ensures that when the stack is
non-empty, the node corresponding to the last write to $\Top$ is the
push with the largest timestamp, i.e.,
\begin{align}
  \label{ref-3b}
  &
    \begin{array}[c]{@{}l@{}l@{}}
      {\it lastVal}_\beta(Top) \neq null \imp{} & \lastPush_\beta \in \unmatchedPush_\alpha  \wedge f ({\it lastVal}_\beta(Top)) = \lastPush_\alpha
    \end{array}
\end{align}
Finally, we have a set of conditions describing the views of threads
over client variables. Together these assertions ensures the client
observation property in \refdef{def:fsim}.
\begin{align}
  \label{ref-5}
  & \forall a\in \mem.\pushedAddr, x \in \GVar_C .\   \tst (\beta.\mview_{f(a)}(x))
       \le \tst (\gamma_C.\tview_t(x))
  \\
  \label{ref-6}
  &
    \begin{array}[c]{@{}l@{}}
      \forall a\in \mem.\pushedAddr, w \in \beta.\writes.\  var(w) = Top \wedge val(w) = a  \imp  \\
      \qquad 
      \forall x \in \GVar_C.\ \tst (\alpha.\mview_{f(a)}(x)) \le \tst(\beta.\mview_w(x))
    \end{array}
  \\
    \label{ref-7}
  & \forall x \in \GVar_C .\ \tst (\gamma_A.\tview_t(x)) \le \tst(\gamma_C.\tview_t(x))
\end{align}

\noindent
We then obtain the following proposition.
\begin{proposition}
  For synchronisation-free clients, $R_{\it TS}$ is a forward simulation
  between the abstract stack object and the Treiber stack.
\end{proposition}

The simulation proof is similar to linearizability proofs in the 
classical (sequentially consistent) setting~\cite{DongolD15}. All
steps except for a successful $\kwcas$ at lines 9 and 16 are
stuttering steps and are not matched with any abstract step. The
successful $\kwcas$ operations corresponds to the a releasing push or
an acquiring pop as allowed by the specification in \refex{ex:stack}. 

\subsection{Mechanisation}
Our deductive proof of the Treiber stack for the RC11-RAR within
Isabelle/HOL has been a significant undertaking. Many properties that
one would take for granted in a sequentially consistent setting are no
longer valid under weak memory. For instance, to guarantee that the
writes to the fields at lines 5 and 8 are seen by the reads in lines
15 and 17, respectively, one must ensure a happens-before relationship
through the releasing $\kwcas$ at line 9 and the acquiring read at
line 13. Moreover, client-side synchronisation guarantees defined by
the specification must also shown to be satisfied.
This verification has been made possible by the proof infrastructure
developed in earlier work~\cite{ECOOP20}, which considered much
simpler examples. Here, we have shown that the proof principles carry
over to pointer-based programs. Additionally, the specifications that
we have developed in \refsec{sec:specifying-c11-style} (and
\refsec{sec:abstr-object-semant}) are appropriate for guaranteeing
client synchronisation through a library object, and our refinement
theory can be used to show that these client-library synchronisation
requirements are also guaranteed by the Treiber stack.

The full development and verification of the Treiber stack is around 12 KLOC
in Isabelle accross 9 different theories. The proof is heavily dependent on
the general proof rules developed for the generalised C11 operational semantics.
A considerable effort was put to identify and prove the local and global
invariants of the stack's push and pop operations. Each of these invariants
was then proved correct locally and globally (interference freedom) in 
the usual Owicki-Gries manner. 
\subsection{Discussion}

The message passing idiom has manifested in our case study in two
different ways. In \reffig{fig:mp-proof} and the corresponding proof
of \refprop{prop:stack-mp} involves message passing through the stack
specification. In the Treiber stack implementation, threads use
message-passing to ensure that the relaxed writes to the ``val'' and
``nxt'' fields of a node within a push are viewed by the pop via
release-acquire synchronisation on the $\Top$ pointer. In both cases,
message passing is succinctly characterised by {\em conditional
  observation} assertion, suggesting that it is a key operator for
reasoning about such behaviours.
Conditional observation assertions capture the idea behind message
passing via ``ownership transfer'' as used in many separation
logics~\cite{DBLP:conf/pldi/TassarottiDV15,DBLP:conf/esop/DokoV17,DBLP:conf/ecoop/KaiserDDLV17,DBLP:journals/sttt/SummersM20,DBLP:conf/tacas/Summers018},
a ``communication specification''~\cite{DBLP:conf/popl/AlglaveC17} and
``protocols''~\cite{DBLP:conf/oopsla/TuronVD14}, which have been used
to ensure happens-before knowledge is passed from one thread to
another. These techniques are however focussed on message passing via
reads and writes \emph{within} a single object and cannot handle
message passing on client threads using library
synchronisation. 


An interesting observation that has been elucidated by our refinement
proof is the fact that the Treiber stack does not need to linearize
its operation in the middle of the abstract operation order as defined
by their timestamps. I.e., each new push / pop operation thereby is
not required to take advantage of the relaxation provided by the
specification in \refex{ex:stack}. This is however not surprising ---
the concrete implementation synchronises all pushes and pops through a
single $\Top$ pointer that is only modified using an RMW
generated by a $\kwcas$.  We conjecture this relaxation would be
required by, for example, a weak memory implementation of the
elimination stack~\cite{DBLP:conf/spaa/HendlerSY04}, which augments a
standard stack with a disjoint elimination mechanism. Operations that
synchronise using the elimination mechanism do not interact with the
main stack data structure, and hence are not guaranteed to advance
thread views. We therefore conjecture that a C11-style relaxed-memory
version of the elimination stack would be an implementation of our
specification, where the stack can be weakly synchronised, allowing it
to take advantage of our relaxed specification. 







%

\section{Related work}

The speed of development in the area has meant that it has become
almost impossible to cover all related works. Below, we provide a
brief overview of different approaches. There are now several
different approaches to program verification that support different
aspects of weak memory using pen-and-paper proofs
(e.g.,~\cite{DBLP:conf/icalp/LahavV15,DBLP:conf/oopsla/TuronVD14,DBLP:conf/popl/AlglaveC17,DBLP:conf/esop/DokoV17}),
model checking
(e.g.,~\cite{DBLP:conf/pldi/Kokologiannakis19,DBLP:conf/pldi/AbdullaAAK19}),
specialised tools
(e.g.,~\cite{DBLP:conf/pldi/TassarottiDV15,DBLP:conf/esop/KrishnaEEJ20,DBLP:conf/esop/SvendsenPDLV18,DBLP:conf/tacas/Summers018}),
and generalist theorem provers (e.g.,~\cite{ECOOP20}). These cover a
variety of (fragments of) memory models and proceed via exhaustive
checking, separation logics, or Hoare-style calculi.

Library abstractions in weak memory have been considered in earlier
works (e.g.,~\cite{DBLP:conf/popl/BattyDG13}), but these were not accompanied
by an associated program logic or verification technique. The use of
abstract methods can be used to ensure weak-memory synchronisation
guarantees has been established in earlier
work~\cite{ifm18,DongolJRA18}, where it has been shown to be necessary
for contextual refinement~\cite{DongolJRA18} and
compositionality~\cite{ifm18}.



Krishna et al. \cite{DBLP:conf/esop/KrishnaEEJ20} have developed an
approach to verifying implementations of weakly consistent
libraries~\cite{DBLP:journals/pacmpl/EmmiE19}. Their techniques are
mechanised using the CIVL verification tool and account for weak
memory relaxations by transitioning over a generic happens-before
relation encoded within a transition system. On the one hand, this
means that their techniques apply to any memory model, but on the
other hand, such a happens-before relation must ultimately be
supplied. Their methods currently cover refinement of libraries, but
do not cover clients that use library specifications. Nevertheless
given that CIVL is aimed at automation, it would be interesting to see
whether the assertion language from earlier works~\cite{ECOOP20} (also
see \refsec{sec:example-verification}) can have a role in further
improving these verification methods.

More recent works include that on {\em robustness} of C11-style
programs, which aims to show ``adequate synchronisation'' so that the
weak memory executions reduce to executions under stronger memory
models~\cite{10.1145/3434285}. Such reductions, although automatic,
are limited to finite state systems, and a small number of
threads. Furthermore, it is currently unclear how they would handle
client-library synchronisation or relaxed non sequentially consistent
specifications. Verification tools for deductive verification of C11-style programs
have been based on separation
logic~\cite{DBLP:conf/tacas/Summers018,DBLP:journals/sttt/SummersM20,DBLP:conf/ecoop/KaiserDDLV17}
and Owicki-Gries
reasoning~\cite{DBLP:journals/corr/abs-2004-02983,ECOOP20}. Those
based on separation logic offer a degree of automation but, as far as
the authors are aware, have not been applied to verification of data
structures. None of these works cover refinement. 

Orthogonal to our paper is a series of works that address more relaxed
memory semantics that allow causality
cycles~\cite{Batty2020,DBLP:conf/popl/KangHLVD17,DBLP:conf/pldi/LeeCPCHLV20,DBLP:journals/pacmpl/JagadeesanJR20},
e.g., the load-buffering litmus test, which our semantics
(like~\cite{DBLP:conf/pldi/LahavVKHD17}) disallows. Although an
important problem, we eschew proofs of correctness in this weaker
memory model since the precise weak memory semantics that
appropriately handles load-buffering, yet rules out thin-air reads is
still a topic of ongoing research. Moreover, we note that many
concurrent object implementations naturally induce intra-thread
dependencies, whereby the relaxed model and the strict model we use
coincide without requiring introduction of any additional
synchronisation. Thus, although weak memory models with causality
cycles are a real phenomenon, there is more work to be done before
they relaxations they offer can be exploited. We consider the problem
of deductive verification under such memory models to be an important
problem for future work.

\section{Conclusions}
\label{sec:conclusion}

In this paper, we have presented a new approach to specifying and
verifying abstract objects over weak memory by extending an existing
operational semantics for RC11 RAR (a fragment of the C11 memory
model). This operational semantics allows one to execute programs in
thread order and accommodates weak memory behaviours via a special
encoding of the state. Moreover, unlike prior works~\cite{ECOOP20}, it
describes how (release-acquire) synchronisation within a library
affects a client state, and vice versa, supporting modular program
development.

Operational semantics are widely acknowledged as being a prerequisite
to developing program logics, complementing declarative approaches
(e.g., \cite{AlglaveMT14,DBLP:journals/pacmpl/RaadDRLV19}), since a
meaning for stepwise program execution is needed to establish meaning
for pre/postconditions. We have demonstrated this by developing an
Owicki-Gries logic for client-library programs, extending prior
work~\cite{DBLP:conf/icalp/LahavV15,ECOOP20}, which did not facilitate
proof modularisation of this nature.

We have also developed a novel operational technique for {\em
  specifying} weak memory concurrent objects, and shown that the
specification allows proofs of client-library synchronisation across
abstract library method calls.  
This has been accompanied by a
refinement theory, that extends standard notions of refinement under
sequential consistency~\cite{DBLP:books/cu/RoeverE1998}.  To exploit
these operational descriptions, we have developed an assertion
language that describes a thread's observations of client-library
states, which is in turn used to verify program invariants and proofs
of refinement.
We have shown that our methodology supports two types of
verification: (1) proofs of correctness of client programs that
\emph{use} abstract libraries and (2) refinement proofs between
abstract libraries and their implementations.
We present the verification of two lock implementations as well as the
Treiber stack adapted for C11. 
This amounts to one of the more sophisticated proofs performed using a
deductive verification approach.

\bibliographystyle{alpha}

\bibliography{references} 
\newpage\appendix
\section{Proof Rules for Clients and Libraries}

Here, we give a set of rules from our Isabelle/HOL development
\cite{Isabelle} that we use in the verification of client-library
programs.

We use $[S]_t^X$, where $S$ is a statement, $t$ is a thread and
$X \in \{L, C\}$ to denote that $S$ is statement of either the library
($L$) or client ($C$) action that is executed by thread $t$. E.g.,
$[x ~ := ~ v]^L_t$ denotes that $x := v$ is a write statement of the
library being executed thread $t$.

We specialise the notion of a covered operation from
\refsec{sec:assertion-language} and define covered writes over shared
locations as follows:
\begin{align*}
  CW[x, u](\sigma) \equiv \forall w .\  w \in \sigma.(W_{|x})^{-1} \land w \not\in \sigma.cvd \imp w = \sigma.\last(W,x) \land \wrval(w) = u
\end{align*}
The assertion $CW[x, u]$ assumes a variable (or location) $x$ and
value $u$, and holds iff all but the last write to $x$ are covered,
and that the value of the last write is $u$. \medskip

%
%



%
%

\begin{enumerate}

  \item $\inference[{}]{ X \in \{L, C\} 
}{\{[x ~ = ~ u]^X_t \} [x ~ := ~ v]^X_t \{[x ~ = ~ u]^X_t\}}$

\vspace{2em}

\item $\inference[{}]{ ~ X \in \{L, C\}
  }{
    \{\exists ~ u ~ . ~ CW[x, u]\}\  [\kwcas^{\sf R}(x, v, ~ v')]^X_t
    \{\exists ~ v ~ . ~ CW[x, v] \}}$
\vspace{2em}

\item $\inference[{}]{ ~ X \in \{L, C\}
}{ \{CW[x, ~ v]^X \}
~~ [r \gets \kwcas^{\sf R}(x, ~ u, ~ u')]_t^X~ \{r \imp [x ~ = ~ u']_t^X\}}$
\vspace{2em}

\item $\inference[{}]{
X \in \{L, C\}
}{ \{\langle x ~ = ~ a\rangle^{L}[y ~ = ~ b]_t^{X} \}~~[a ~ \leftarrow^A ~ x]_t^L~ \{[y ~ = ~ b]_t^X\} ~
 ~  ~ }$
\vspace{2em}

\item $\inference[{}]{ ~ X, X' \in \{L, C\} }{ \{[x ~ =_t ~ m]^X\}
~~ [r \gets \kwcas^{\sf R}(x, ~ u, ~ u')]_t^{X'}~ \{\neg r \imp [x ~ = ~ m]_t^X\}}$
\vspace{2em}

\item $\inference[{}]{X\in\{L, C\} \qquad t ~ \neq ~  ~ t'
 \qquad x ~ \neq ~  ~ y
}{ ~ \{\neg ~ [x~ \approx ~ u]_{t'}^X 
~~\land~~[y ~ = ~ v]_t^X\} [r \gets \kwcas^{\sf R}(x, ~ l, ~ u)]_t^X \{r \imp \langle x ~ = ~ u \rangle^{X} [y ~ = ~ v]_{t'}^{X}\}
 ~  ~ }$
\vspace{2em}

\item $\inference[{}]{
X\in\{L, C\} \qquad k ~ \neq ~  ~ u
\qquad t ~ \neq ~  ~ t' \qquad x ~ \neq ~  ~ y
}{ ~\{\langle x ~ = ~ u \rangle^{X} [y ~ = ~ v]_{t'}^{X}\} [r \gets \kwcas^{\sf R}(x, ~ l, ~ k)]_t^{X} \{r \imp \langle x ~ = ~ u \rangle^{X} [y ~ = ~ v]_{t'}^{X}\}
 ~  ~ }$
\vspace{2em}

\item $\inference[{}]{X\in\{L, C\} \qquad  x\neq  y }{\{[y ~ = ~ v]_t^{X} ~ ~~\land~~ ~ CW[x, ~ m]^{X}\} [r \gets \kwcas^{\sf R}(x, ~ m, ~ u)]_t^{X} \{r \imp \langle x ~ = ~ u\rangle^{X}[y ~ = ~ v ~]_t^{X}\}}$
\vspace{2em}

\item $\inference[{}]{X \in \{L, C\}~~
}{\{true\} [v ~ \leftarrow^{[A]} ~ x]_t^{X} \{[x ~ \approx ~ v]_t^{X}\} ~ 
 ~  ~ }$
\vspace{2em}

\item $\inference[{}]{
X \in \{L, C\}
}{ ~\{CW[x, ~ v]^{X}\}~~ [r \gets \kwcas^{\sf R}(x, ~ v, ~ v')]_t^{X} ~~ \{r \imp CW[x, ~ v']^{X}\} ~ 
 ~  ~ }$
\vspace{2em}

\item $\inference[{}]{
X \in \{L, C\}
}{\{CW[x, ~ v]^{X}\} [r \gets \kwfai(x)]^{X}_t~ \{CW[x, ~ (v+1)]^{X}\} ~ 
 ~  ~ }$
\vspace{2em}

\item $\inference[{}]{
X \in \{L, C\}
}{ \{True\}~~ [r \gets \kwcas^{\sf R}(x, ~ v, ~ v')]_t^{X}~ \{r \imp [x ~ \approx ~ v']_t^{X}\} ~ 
 ~  ~ }$
\vspace{2em}

\item $\inference[{}]{
 X \in \{L, C\} 
~~\land~~x ~ \neq ~  ~ y
}{\{\neg ~ [x ~ \approx ~ u]_t^{X}\} ~ [r \gets \kwcas^{\sf R}(y, ~ v, ~ v')]_{t'}^{X} ~ \{\neg ~ [x ~ \approx ~ u]_t^{X}\}}$
\vspace{2em}

\item $\inference[{}]{
X \in \{L, C\}
}{\{\neg ~ [x ~ \approx ~ u]_t^{X}\} ~[v ~ \leftarrow ~ y]_{t'}^{X} ~ \{\neg ~ [x ~ \approx ~ u]_t^{X}\}}$
\vspace{2em}

\item $\inference[{}]{
 X \in \{L, C\} \qquad y ~ \neq ~  ~ x
}{\{\langle x ~ = ~ v \rangle^{X}[x ~ = ~ v]^{X}_t \} ~[v \gets \kwfai(y)]_{t'}^{X}~ \{\langle x ~ = ~ v \rangle^{X}[x ~ = ~ v]^{X}_t\}}$
\vspace{2em}

\item $\inference[{}]{X \in \{L, C\}
}{ ~\{\langle x ~ = ~ v \rangle^{X}[x ~ = ~ v]^{X}_t\} ~~ [v ~ \leftarrow^A ~ x]_t^{X} ~~ \{[x ~ = ~ v]_t^{X}\} ~ 
 ~  ~ }$
\vspace{2em}

\item $\inference[{}]{
X \in \{L, C\} \qquad t ~ \neq ~  ~ t'
}{\{\neg ~ [x ~ \approx ~ u]_t^{X} ~ 
~~\land~~[x ~ = ~ v]_{t'}^{X}\} ~[x ~ :=^{\sf R}~ u]_{t'}^{X}~ \{\langle x ~ = ~ u\rangle^{X}[x ~ = ~ u]_t^{X}\} ~ 
 ~  ~ }$
\vspace{2em}

\item $\inference[{}]{
X \in \{L, C\}
\qquad x ~ \neq ~  ~ y
}{\{\langle x ~ = ~ u\rangle^{X}[x ~ = ~ u]_t^{X}\} ~[y ~ := ~ v]_{t'}^{X}~ \{\langle x ~ = ~ u\rangle^{X}[x ~ = ~ u]_t^{X}\} ~ 
 ~  ~ }$
\vspace{2em}

\end{enumerate}

\section{Hoare Logic Rules for The Abstract Stack Object}

In this section we provide a set of Hoare logic proof rules that we
used in our Isabelle/HOL development to prove programs that use the
abstract specification of the stack object. The rules relate
observation assertions for abstract method calls.

Here we assume that the
pop operations are part of the library, thus we omit the superscript
$L$ on the assertions and operations related to the stack. Recall that
$m^{[\sf Z]}(v)_t$ denotes that the annotation
${\sf Z} \in \{{\sf R}, {\sf A}\}$ is optional for the method $m$.

\begin{enumerate}

\item $\inference[{}]{}{\{[pop(u)]_t\}~~ push^{[\sf R]}(v)_t  ~~\{[pop(v)]_t\} ~ 
 ~  ~ }$\vspace{2em}

\item $\inference[{}]{}{\{[pop(u)]_t\}~~ push^{[\sf R]}(v)_t  ~~\{[pop(v)]_t\} ~ 
 ~  ~ }$\vspace{2em}

\item $\inference[{}]{
}{\{[pop(u)]_t\} pop^{[\sf A]}(v)_t \{v ~ = ~ u\}
 ~  ~ }$\vspace{2em}

\item $\inference[{}]{u \neq empty}{
 \{\neg ~ \langle pop(u) \rangle_{t'}\wedge [y ~ =_t ~ v]^C\}\ push^{\sf R}(u)_t\
 \{ \langle pop(u) \rangle [y ~ = ~ v]_{t'}\}}$\vspace{2em}

\item $\inference[{}]
  {
    u \neq empty}{\{\langle pop(u)\rangle [y ~ = ~ v]_t^C\} pop^{\sf A}(u)_t \{[y ~ = ~ v]_t\}}$\vspace{2em}

\item $\inference[{}]{
}{\{[x ~ = ~ v]_t^C\}\ push^{\sf [R]}(u)_{t'} \ \{[x ~ = ~ v]_t^C\}}$\vspace{2em}

\item $\inference[{}]{
}{\{[x ~ = ~ v]_{t'}^C \}\ pop^{[\sf A]}(u)_t \ \{[x ~ = ~ v]_{t'}^C\}}$\vspace{2em}

\item $\inference[{}]{
}{\{\neg ~ \langle pop(u)\rangle_{t'}\}\ [x := v]_t^C\ \{\neg ~ \langle pop(u)\rangle_{t'}\}
 ~  ~ }$\vspace{2em}

\item $\inference[{}]{
}{\{\neg ~ \langle pop(u)\rangle_{t'}\}\ [v ~ \leftarrow ~ x]_t^C\ \{\neg ~ \langle pop(u)\rangle_{t'}\}}$\vspace{2em}

\item $\inference[{}]{u \neq  empty}{\{\neg ~ \langle pop(u)\rangle_{t'}\} \ pop^{[A]}(v)_t \ \{\neg ~ \langle pop(u)\rangle_{t'}\}}$\vspace{2em}

\item $\inference[{}]{
u \neq z
\qquad u \neq empty 
\qquad z \neq empty
}{\{\langle pop(u)\rangle [y ~ = ~ v]_t^C\}\ pop^{[A]}(z)_t \ \{\langle pop(u)\rangle [y ~ = ~ v]_t^C\}}$\vspace{2em}

\item $\inference[{}]{u\neq  z
~~ \qquad u \neq empty
}{\{\langle pop(u)\rangle [y ~ = ~ v]_t^C\}\ pop^{[A]}(z)_t \ \{\langle pop(u)\rangle [y ~ = ~ v]_t^C\}}$\vspace{2em}


\item $\inference[{}]{}{\{\neg ~ \langle pop(u)\rangle_{t}\}\ [x ~ := ~ z]_{t'}^C \ \{\langle pop(u)\rangle [x ~ = ~ v]_t^C\}}$\vspace{2em}

\end{enumerate}

\end{document}